\documentclass[zpreprint,zbstdefault]{zeus_paper}
\usepackage[english]{babel}
\chardef\usc=95
\chardef\til=126
\catcode`\@=11 
\DeclareRobustCommand\xdotspace{\futurelet\@let@token\@xdotspace}
\def\@xdotspace{%
  \ifx\@let@token.\else
  \ifx\@let@token\bgroup.\else
  \ifx\@let@token\egroup.\else
  \ifx\@let@token\/.\else
  \ifx\@let@token\ .\else
  \ifx\@let@token~.\else
  \ifx\@let@token!.\else
  \ifx\@let@token,.\else
  \ifx\@let@token:.\else
  \ifx\@let@token;.\else
  \ifx\@let@token?.\else
  \ifx\@let@token/.\else
  \ifx\@let@token'.\else
  \ifx\@let@token).\else
  \ifx\@let@token-.\else
  \ifx\@let@token\@xobeysp.\else
  \ifx\@let@token\space.\else
  \ifx\@let@token\@sptoken.\else
   .\space
   \fi\fi\fi\fi\fi\fi\fi\fi\fi\fi\fi\fi\fi\fi\fi\fi\fi\fi}
\catcode`\@=12 

\newcommand{\stru}[2]{%
   \relax\ifmmode\hbox{\vrule height#1 depth#2 width0pt}%
   \else\vrule height#1 depth#2 width0pt\fi}

\newcommand{\Ronum}[1]{\uppercase\expandafter{\romannumeral#1}}
\newcommand{\ronum}[1]{\expandafter{\romannumeral#1}}
\DeclareRobustCommand{\LaTeXZ}{%
  \LaTeX\kern-.05em4\kern-.1em
  {\raisebox{-0.2ex}{$\scriptstyle\text{ZEUS}$}}\xspace}

\DeclareMathAlphabet{\mathbf}{OT1}{cmr}{bx}{sl}
\newcommand{\eVdist}{\kern-0.06667em}

\newcommand{\mev}{{\,\text{Me}\eVdist\text{V\/}}}
\newcommand{\gev}{{\,\text{Ge}\eVdist\text{V\/}}}

\newcommand{\pbi}{\,\text{pb}^{-1}}

\newcommand{\Tesla}{\,\text{T}}

\newcommand{\slashfrac}[2]{%
  \raisebox{0.5ex}{\ensuremath #1}\kern-0.12em/\kern-0.08em
  \raisebox{-.8ex}{\ensuremath #2}}

\newcommand{\sqr}[3]{%
    {\vcenter{\hrule height.#3ex\hbox{\vrule width.#2ex height#1ex
     \kern#1ex\vrule width.#3ex}\hrule height.#2ex}}}

\catcode`\@=11 
\newcommand{\parenbar}{\mathpalette\p@renb@r}
\def\p@renb@r#1#2{\vbox{%
  \ifx#1\scriptscriptstyle \dimen@.7em\dimen@ii.2em\else
  \ifx#1\scriptstyle \dimen@.8em\dimen@ii.25em\else
  \dimen@1em\dimen@ii.4em\fi\fi \offinterlineskip
  \ialign{\hfill##\hfill\cr
    \vbox{\hrule width\dimen@ii}\cr
    \noalign{\vskip-.3ex}%
    \hbox to\dimen@{$\mathchar300\hfil\mathchar301$}\cr
    \noalign{\vskip-.3ex}%
    $#1#2$\cr}}}
\catcode`\@=12 

\newcommand{\IP}{{\rm I$\kern-0.01667em$P}\xspace}

\mathchardef\qsm=63
\mathchardef\pls=43
\mathchardef\mns=512
\mathchardef\plm=518
\mathchardef\eql=61
\mathchardef\smallleft=300
\mathchardef\smallright=301
\mathchardef\les=316
\mathchardef\gre=318
\mathchardef\leq=532
\mathchardef\grq=533

\catcode`\@=11 
\newcounter{pict@width}
\newcounter{pict@height}
\newlength{\pict@scale}
\newcommand{\psfigadd}[4]{%
\setcounter{pict@width}{1*\ratio{#2+\pict@scale/2}{\pict@scale}}
\setcounter{pict@height}{1*\ratio{#3+\pict@scale/2}{\pict@scale}}
\setlength{\unitlength}{\pict@scale}
\hbox to #2{\hspace{-\fill}\begin{picture}(\thepict@width,\thepict@height)
\put(0,0){\psfig{figure=#1,width=#2,height=#3,clip=}}
\SetScale{0.283466457}
\SetWidth{1.763889}
{#4}
\end{picture}}
}
\newcounter{pict@widthfst}
\newcounter{pict@widthscd}
\newcounter{pict@widthtot}
\newcommand{\psfigaddtwo}[7]{%
\setcounter{pict@widthfst}{1*\ratio{#2+\pict@scale/2}{\pict@scale}}
\setcounter{pict@widthscd}{1*\ratio{#2+#4+\pict@scale/2}{\pict@scale}}
\setcounter{pict@widthtot}{1*\ratio{#2+#4+#6+\pict@scale/2}{\pict@scale}}
\setcounter{pict@height}{1*\ratio{#3+\pict@scale/2}{\pict@scale}}
\setlength{\unitlength}{\pict@scale}
\hbox{\hspace{-\fill}\begin{picture}(\thepict@widthtot,\thepict@height)
\put(0,0){\psfig{figure=#1,width=#2,height=#3,clip=}}
\put(\thepict@widthscd,0){\psfig{figure=#5,width=#6,height=#3,clip=}}
\SetScale{0.283466457}
\SetWidth{1.763889}
{#7}
\end{picture}}
}
\newcommand{\psfigror}[4]{%
\setcounter{pict@width}{1*\ratio{#2+\pict@scale/2}{\pict@scale}}
\setcounter{pict@height}{1*\ratio{#3+\pict@scale/2}{\pict@scale}}
\setlength{\unitlength}{\pict@scale}
\hbox{\begin{picture}(\thepict@width,\thepict@height)
\put(0,\thepict@height){\psfig{figure=#1,width=#3,height=#2,clip=,angle=270}}
\SetScale{0.283466457}
\SetWidth{1.763889}
{#4}
\end{picture}}
}
\newcommand{\psfigrol}[4]{%
\setcounter{pict@width}{1*\ratio{#2+\pict@scale/2}{\pict@scale}}
\setcounter{pict@height}{1*\ratio{#3+\pict@scale/2}{\pict@scale}}
\setlength{\unitlength}{\pict@scale}
\hbox{\begin{picture}(\thepict@width,\thepict@height)
\put(0,0){\psfig{figure=#1,width=#3,height=#2,clip=,angle=90}}
\SetScale{0.283466457}
\SetWidth{1.763889}
{#4}
\end{picture}}
}
\catcode`\@=12 
\newlength\listtextwidth

\catcode`\@=11 
\newlength{\@tabfninsert}
\newlength{\@tabfnwidth}
\newcommand{\tabfootnote}[2]{%
  \setlength{\@tabfninsert}{0.8em}
  \setlength{\@tabfnwidth}{\textwidth}
  \addtolength{\@tabfnwidth}{-\@tabfninsert}
  \addtolength{\@tabfnwidth}{-0.4em}
  \noindent\makebox[\@tabfninsert][r]{\footnotesize$^{#1}$\hfil}\hfill%
  \parbox[t]{\@tabfnwidth}{\footnotesize #2\hfill}}
\catcode`\@=12 

\newcommand{\xpom}{x_{I\hspace{-0.8mm}P}}
\newcommand{\xgam}{x_{\gamma}^{obs}}
\newcommand{\zpom}{z_{I\hspace{-0.8mm}P}^{obs}}

\renewcommand{\xgam}{x_{\gamma}^{\rm{obs}}}
\renewcommand{\xpom}{x_{I\!P}}
\renewcommand{\zpom}{z_{I\!P}^{\rm{obs}}}
\newcommand{\coll}{Collaboration}
\def\citeH1Fit2{{\cite{%
zfp:c76:613%
}}\xspace}
\def\citeH12002Fit{{\cite{%
h1:2002fit:eps03%
}}\xspace}
\def\citeDiffDISH1paper{{\cite{%
epj:c20:29%
}}\xspace}
\def\citeDiffDISH1conf{{\cite{%
h1:diffdis:ichep04%
}}\xspace}
\def\citeJETSET1{{\cite{%
cpc:46:43%
}}\xspace}
\def\citeJETSET2{{\cite{%
cpc:82:74%
}}\xspace}
\def\citeCAL{{\cite{%
nim:a309:77,*nim:a309:101,*nim:a321:356,*nim:a336:23%
}}\xspace}

\def\citeLUMI1{{\cite{%
desy-92-066%
}}\xspace}
\def\citeLUMI2{{\cite{%
zfp:c63:391%
}}\xspace}
\def\citeLUMI3{{\cite{%
acpp:b32:2025%
}}\xspace}
\def\citeInclDijetPHP1{{\cite{%
epj:c11:35%
}}\xspace}
\def\citeInclDijetPHP2{{\cite{%
epj:c23:615%
}}\xspace}
\begin{document}
\prepnum{DESY--07--161}
\title{\boldmath
Diffractive photoproduction of dijets \\
in $ep$ collisions at HERA
}                                                       
                    
\author{ZEUS Collaboration}
\date{25th September 2007}

\abstract{
Diffractive photoproduction of dijets was measured with the ZEUS detector 
at the $ep$ collider HERA using an integrated luminosity of $77.2\pbi$.   
The measurements were made in the kinematic range $Q^2 < 1\gev^2$,  
$0.20 < y < 0.85$ and $\xpom < 0.025$, where 
$Q^2$ is the photon virtuality, $y$ is the inelasticity  
and $\xpom$ is the fraction of the proton momentum taken by 
the diffractive exchange.
The two jets with the highest transverse energy, $E_T^{\rm{jet}}$, 
were required
to satisfy $E_T^{\rm{jet}} > 7.5$ and $6.5\gev$, respectively, and to lie in 
the pseudorapidity range $-1.5 < \eta^{\rm{jet}} < 1.5$. 
Differential cross sections 
were compared to  perturbative QCD calculations using available 
parameterisations of diffractive parton distributions  of the proton.  
}

\makezeustitle

\def\3{\ss}                                                                                        
\pagenumbering{Roman}                                                                              
\begin{center}                                                                                     
{                      \Large  The ZEUS Collaboration              }                               
\end{center}                                                                                       
  S.~Chekanov$^{   1}$,                                                                            
  M.~Derrick,                                                                                      
  S.~Magill,                                                                                       
  B.~Musgrave,                                                                                     
  D.~Nicholass$^{   2}$,                                                                           
  \mbox{J.~Repond},                                                                                
  R.~Yoshida\\                                                                                     
 {\it Argonne National Laboratory, Argonne, Illinois 60439-4815, USA}~$^{n}$                       
\par \filbreak                                                                                     
  M.C.K.~Mattingly \\                                                                              
 {\it Andrews University, Berrien Springs, Michigan 49104-0380, USA}                               
\par \filbreak                                                                                     
  M.~Jechow, N.~Pavel~$^{\dagger}$, A.G.~Yag\"ues Molina \\                                        
  {\it Institut f\"ur Physik der Humboldt-Universit\"at zu Berlin,                                 
           Berlin, Germany}~$^{b}$
% ...
\newpage  

\enlargethispage{5cm}
$^{\    1}$ supported by DESY, Germany \\                                                          
$^{\    2}$ also affiliated with University College London, UK \\                                  
% ...
$^{\    5}$ retired \\                                                                             
% ...
$^{\dagger}$ deceased
%
\newpage   
                                                           %
                                                           %
\begin{tabular}[h]{rp{14cm}}                                                                                                        
$^{a}$ &  supported by the Natural Sciences and Engineering Research Council of Canada (NSERC) \\  
$^{b}$ &  supported by the German Federal Ministry for Education and Research (BMBF), under        
          contract numbers 05 HZ6PDA, 05 HZ6GUA, 05 HZ6VFA and 05 HZ4KHA\\                         
$^{c}$ &  supported in part by the MINERVA Gesellschaft f\"ur Forschung GmbH, the Israel Science   
          Foundation (grant no. 293/02-11.2) and the U.S.-Israel Binational Science Foundation \\  
$^{d}$ &  supported by the German-Israeli Foundation and the Israel Science Foundation\\           
$^{e}$ &  supported by the Italian National Institute for Nuclear Physics (INFN) \\                
$^{f}$ &  supported by the Japanese Ministry of Education, Culture, Sports, Science and Technology 
          (MEXT) and its grants for Scientific Research\\                                          
$^{g}$ &  supported by the Korean Ministry of Education and Korea Science and Engineering          
          Foundation\\                                                                             
$^{h}$ &  supported by the Netherlands Foundation for Research on Matter (FOM)\\                   
$^{i}$ &  supported by the Polish State Committee for Scientific Research, grant no.               
          620/E-77/SPB/DESY/P-03/DZ 117/2003-2005 and grant no. 1P03B07427/2004-2006\\             
$^{j}$ &  partially supported by the German Federal Ministry for Education and Research (BMBF)\\   
$^{k}$ &  supported by RF Presidential grant N 8122.2006.2 for the leading                         
          scientific schools and by the Russian Ministry of Education and Science through its grant
          Research on High Energy Physics\\                                                        
$^{l}$ &  supported by the Spanish Ministry of Education and Science through funds provided by     
          CICYT\\                                                                                  
$^{m}$ &  supported by the Particle Physics and Astronomy Research Council, UK\\                   
$^{n}$ &  supported by the US Department of Energy\\                                               
$^{o}$ &  supported by the US National Science Foundation. Any opinion,                            
findings and conclusions or recommendations expressed in this material                             
are those of the authors and do not necessarily reflect the views of the                           
National Science Foundation.\\                                                                     
$^{p}$ &  supported by the Polish Ministry of Science and Higher Education                         
as a scientific project (2006-2008)\\                                                              
$^{q}$ &  supported by FNRS and its associated funds (IISN and FRIA) and by an Inter-University    
          Attraction Poles Programme subsidised by the Belgian Federal Science Policy Office\\     
$^{r}$ &  supported by the Malaysian Ministry of Science, Technology and                           
Innovation/Akademi Sains Malaysia grant SAGA 66-02-03-0048\\                                       
\end{tabular}                                                                                      
                                                           %

\clearpage
\pagenumbering{arabic} 
\pagestyle{plain}
\section{Introduction}
\label{sec-int}

In diffractive electron-proton scattering, the proton loses a small 
fraction of its energy and either emerges from the scattering intact, 
$e p \rightarrow e X p$, or dissociates into a low-mass state $N$, 
$e p \rightarrow e X N$. A large rapidity gap (LRG) separates the 
hadronic system $X$ with invariant mass $M_{X}$ and the final-state 
proton $p$.

In the framework of Regge phenomenology  
\cite{lit-regge,*lit-D1b,*lit-D1c}, 
diffractive interactions are ascribed to the exchange of a trajectory 
with vacuum quantum numbers, the Pomeron trajectory. In quantum 
chromodynamics (QCD), the diffractive factorisation theorem 
\cite{pl:b323:201,*lit-D3,lit-QCDfact}
states that the diffractive cross section in deep inleastic scattering (DIS) 
can be expressed as the convolution of universal partonic cross sections and 
a specific type of parton distribution functions (PDF), the diffractive PDF 
(dPDF). Diffractive PDFs are interpreted as the number density of partons 
conditional on the observation of a  
 diffracted proton in the final state. 
The dPDFs \cite{epj:c38:43,lit-D6,lit-groys,lit-D9}
have been determined from the HERA inclusive measurements of the diffractive 
structure function $F_2^D$ \cite{epj:c38:43,lit-D6}, defined in analogy with 
the proton structure function $F_2$,  
and were used  for  input to calculations of hard  diffractive 
processes at HERA, Tevatron and LHC 
\cite{lit-Dstar,lit-H1recent,lit-ddis2,lit-ddis1,lit-LHC3,*lit-LHC1,*lit-LHC4,*lit-LHC2,lit-tevatron}. 

\begin{figure}[p]
  \begin{center}
    \includegraphics[width=0.4\textwidth]{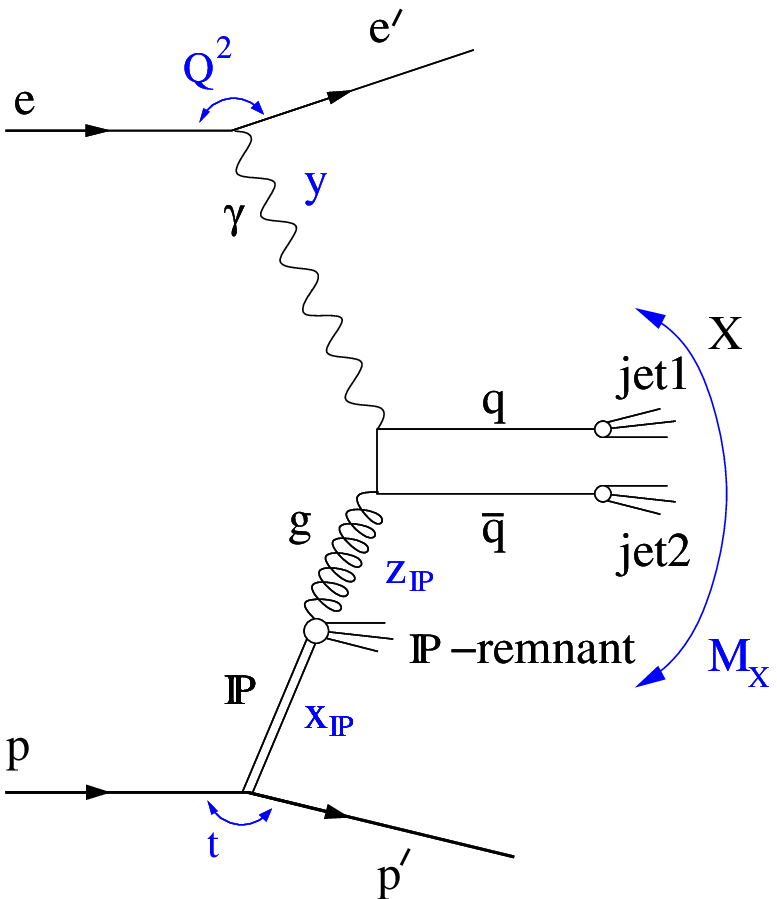} \hskip1.5cm
    \includegraphics[width=0.4\textwidth]{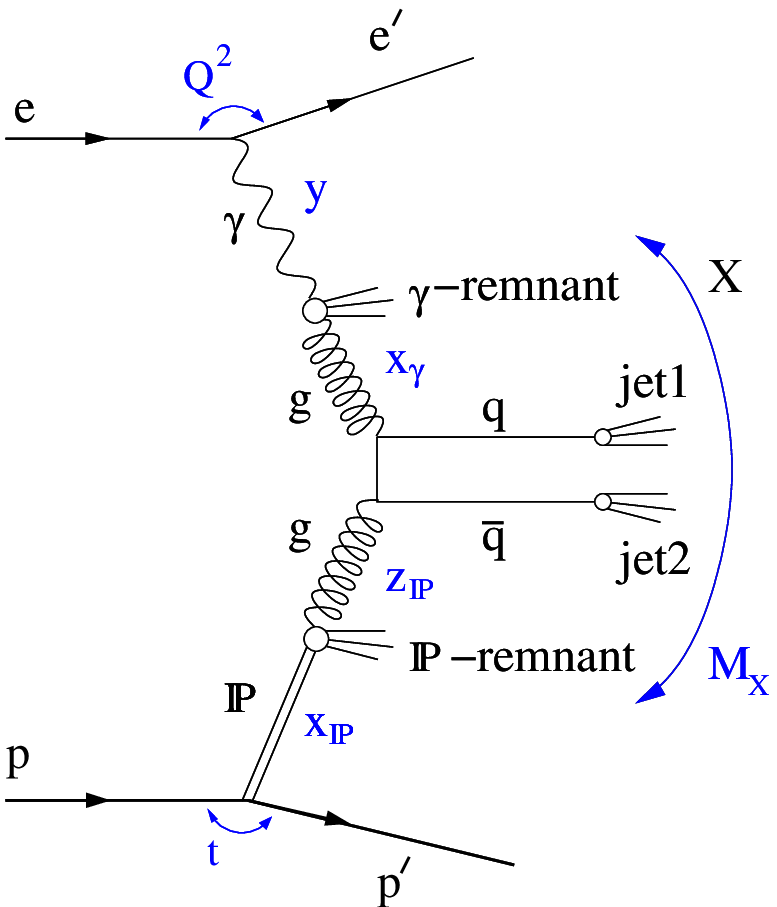} \\
    {\bf  (a) \hskip7.3cm (b) \hspace*{2.3cm} }
  \end{center}
   \caption{Leading-order diagrams for {\sl (a)}\/ direct and {\sl (b)}\/  
     resolved processes in diffractive photoproduction of dijets at HERA.   
       The variables shown in the plots are described in
     the text.}
   \label{fig-diagrams}
   \vfill
\end{figure}

Diffractive collisions producing a state  $X$ with a dijet system 
are a particularly interesting component of diffractive $e p$ interactions.
The transverse energies of the jets provide a hard scale, ensuring the 
applicability of perturbative QCD at the small photon virtualities
considered here.
 In photoproduction, at leading order (LO) of QCD, two 
types of  processes  contribute to dijet photoproduction, namely  direct and 
resolved photon processes. In direct photon processes, the exchanged 
photon participates as a point-like particle,  interacting with a 
gluon from the incoming proton (photon-gluon fusion, 
Fig.~\ref{fig-diagrams}a). Thus, these processes are  directly 
sensitive to the gluon content of the diffractive exchange. 
In resolved photon processes, the photon behaves as a source of partons, 
one of which interacts with a parton from the diffractive exchange 
(Fig.~\ref{fig-diagrams}b). For resolved photon  processes, which resemble 
hadron-hadron interactions, QCD factorisation is not expected to hold 
\cite{lit-QCDfact,lit-QCDfact2,*lit-QCDfact3}.
Further interactions between partons from the 
photon and the proton may fill the rapidity gap, leading to a 
breakdown of hard-scattering factorisation and causing a suppression of 
the diffractive photoproduction cross section. Such a mechanism 
was proposed to explain the suppression of the measured cross sections 
for hard diffractive hadron-hadron scattering at the Tevatron with respect 
to expectations based on dPDFs obtained at HERA \cite{prl:84:5043}. 
For the diffractive resolved photoproduction, an eikonal 
model \cite{lit-D14,*lit-D15} 
predicts a cross-section suppression by about a factor of three. In the 
framework of another model \cite{lit-bialas}, assuming that 
diffractive collisions reflect the absorption of an 
incident particle-wave, it
 has been argued that the strong factorisation breaking  observed in 
diffractive hadron-induced processes should  not be 
seen in photon-induced processes. 

This analysis presents measurements of the diffractive photoproduction 
of dijets using the ZEUS detector at HERA. 
A 30-fold increase in luminosity was achieved compared to the previous 
ZEUS analysis 
\cite{lit-previous}.  This, in combination with the addition 
of a new forward\footnote{The ZEUS coordinate system is a right-handed 
Cartesian system, with the $Z$ axis pointing in the proton beam direction, 
referred to as the ``forward direction'', and the $X$ axis pointing left 
towards the centre of HERA.  The coordinate origin is at the nominal 
interaction point.\xspace} detector, allows  measurements to be made in 
a wider kinematic range. Differential cross sections based on these 
measurements are compared to next-to-leading-order (NLO) QCD 
predictions at the hadron level. 
The comparisons are also made separately for  subsamples enriched with 
direct and resolved photoproduction.
A similar study has been recently published by the H1 Collaboration 
\cite{lit-H1recent}.

\section{Experimental set-up}
\label{sec-exp}

This measurement is based on the data taken with the ZEUS detector at 
the $ e p$ collider HERA in 1999-2000 when electrons or positrons of
$27.5\gev$ were collided with protons of $920\gev$. The sample used for 
this study corresponds to an integrated luminosity $\mathcal{L}=77.2\pbi$
($12.1\pbi$ and $65.1\pbi$ for the $e^{-} p$ and $e^{+} p$ 
samples, respectively)\footnote{From now on, the word ``electron'' will be 
used as a generic term for both electron and positron.\xspace}.
A detailed description of the ZEUS detector can be 
found elsewhere \cite{lit-ZEUSdet}.
A brief outline of the components that are most relevant for this analysis 
is given below. 

Charged particles are tracked in the central tracking detector (CTD) 
\cite{nim:a279:290,*npps:b32:181,*nim:a338:254}, which operates in a 
magnetic field of $1.43\Tesla$ provided by a thin superconducting 
coil.  The CTD consists of $72$ cylindrical drift chamber layers, 
organised in $9$ superlayers covering the polar-angle region 
$15^{\circ} < \theta < 164^{\circ}$.  The transverse-momentum 
resolution for full-length tracks is 
$\sigma (p_T)/p_T = 0.0058p_T \oplus 0.0065 \oplus 0.0014/p_T$, 
with $p_T$ in $\gev$.

The high-resolution uranium-scintillator calorimeter (CAL) \citeCAL 
consists of three parts: 
the forward (FCAL), the barrel (BCAL) and the rear (RCAL) calorimeters.
Each part is subdivided transversely into towers and longitudinally into 
one electromagnetic section (EMC) and either one (in RCAL) or two 
(in BCAL and FCAL) hadronic sections (HAC). The smallest subdivision 
of the calorimeter is called a cell. The CAL energy resolutions, 
as measured under test beam conditions, are $\sigma(E)/E=0.18/\sqrt{E}$ 
for electrons and $\sigma(E)/E=0.35/\sqrt{E}$ for hadrons, with $E$ in $\gev$.

In 1998, the forward plug calorimeter (FPC) 
\cite{lit-FPC}
was installed in the $20 \times 20$ cm$^2$ beam hole of the FCAL, with 
a small hole of radius 3.15 cm in the centre to accommodate the beam 
pipe. The FPC increased the forward calorimetric coverage by about one 
unit in pseudorapidity to $\eta\lesssim 5$. 

The backing calorimeter (BAC) 
 consists of 
proportional tube chambers placed in the gap of the iron yoke. In the 
present analysis it was used in conjunction with the CTD and the CAL 
to identify  cosmic muons that traversed the yoke.

The luminosity was measured from the rate of the bremsstrahlung process 
$e p \rightarrow e \gamma p$. The resulting small-angle energetic photons 
were  measured by the luminosity monitor 
\cite{desy-92-066,*zfp:c63:391,*acpp:b32:2025}, 
a lead-scintillator calorimeter placed in the HERA tunnel at $Z = -107$ m.

\section{Kinematics and reconstruction of variables}
\label{sec-Theory}

Diffractive photoproduction in $e p$ scattering (Fig.~\ref{fig-diagrams}),
\begin{center}
$e(e) + p(p) \rightarrow e(e^\prime) + X(X) + p(p^\prime)$, 
\end{center}
is described in terms of the four-momenta of the incoming and scattered 
electrons, $e$ and $e^{\prime}$, of the incoming and  scattered protons, 
$p$ and $p^{\prime}$, and  of the hadronic system, $X$. The following 
kinematic variables are defined: 
the photon virtuality, $Q^2=-q^2$, where $q=e -e^\prime$, the squared 
photon-proton centre-of-mass energy, $W^2 = (p + q)^2$, and the fraction 
of the electron energy transferred to the proton in its rest frame 
(inelasticity),

\begin{equation*}
y = \frac{p \cdot q}{p \cdot e} \simeq \frac{W^2}{2 p \cdot e} \enspace .
\end{equation*}

The reaction can be considered to proceed through the interaction of the 
virtual photon with the diffractive exchange (Pomeron, $I\hspace{-0.9mm}P$).
This process 
is described by the invariant mass, $M_{X}$, of the hadronic system $X$ 
and the fraction of the proton momentum carried by the diffractive exchange

\begin{equation*}
x_{I\hspace{-0.8mm}P} = \frac{(p - p^{\prime}) \cdot q} {p \cdot q} \enspace .
\end{equation*}

In the present data, the state $X$ contains a dijet system as the result 
of a hard scattering process. The partons from the resolved photon and 
the diffractive exchange participating in the interaction have fractional 
momenta given by  

\begin{equation*}
x_{\gamma} = \frac{p \cdot u} {p \cdot q} \enspace , 
\end{equation*}
where $u$ is the four-momentum of the parton in the resolved photon, and 

\begin{equation*}
z_{I\hspace{-0.8mm}P} = \frac{v \cdot q}{(p-p^{\prime}) \cdot q} \enspace ,
\end{equation*}
where $v$ is the four-momentum of the parton in the diffractive exchange.
 
Energy flow objects (EFOs) were reconstructed from CAL clusters and CTD 
tracks and combine the CTD and CAL information to optimise the resolution 
of the reconstructed kinematic variables 
\cite{lit-briskin,*epj:c11:35}. 
The EFOs were additionally corrected for energy loss due to inactive material 
in front of the CAL  
\cite{lit-turcato}.

The mass $M_X$ of the hadronic system $X$ was reconstructed as
\begin{center}
$M_X = \sqrt{\sum_h{(E - p_Z)_h  \cdot \sum_h (E + p_Z)_h}} \enspace ,$
\end{center}
where the index $h$ runs over all EFOs. The quantities $E$ and $p_Z$ 
indicate the energy and the longitudinal momentum of the EFOs, respectively. 

The  inelasticity, $y$, was estimated from the EFOs 
according to the Jacquet-Blondel method 
\cite{lit-jb} as
\begin{center}
$y_{\rm{J\!B}} = \sum_h {(E-p_Z)}_h / 2E_e \enspace ,$
\end{center}
where $E_e$ is the initial electron energy. For events with an electron 
candidate, 
the inelasticity was also determined from the scattered electron, $y_e$.

The longitudinal momentum fraction transferred from the proton to the 
diffractive exchange, $\xpom$, was reconstructed as 
\begin{center}
$\xpom = \sum_h (E+p_Z)_h / 2E_p \enspace $,
\end{center}
where  $E_p$ is the initial proton energy.

The jets were reconstructed from the EFOs by using the $k_T$ algorithm  
\cite{lit-ktalgorithm} in the longitudinally inclusive mode \cite{pr:d48:3160} 
in the laboratory frame. The variables $E_T^{\rm{jet1,2}}$ and 
$\eta^{\rm{jet}1,2}$ characterise the two jets with  highest transverse 
energy, $E_T$, with $E_T^{\rm{jet1}} > E_T^{\rm{jet2}}$.   
For the variables $x_{\gamma}$ and $z_{I\hspace{-0.8mm}P}$, which  
are not measurable directly, the observable estimators $\xgam$ 
\cite{lit-xgamma} and $\zpom$ were reconstructed as 

\begin{eqnarray*}
\xgam &=& \frac{\sum_{\rm{jet}1,2} E_T^{\rm{jet}} e^{-\eta^{\rm{jet}}}}{2 y E_e}\enspace , \\
\zpom &=& \frac{\sum_{\rm{jet}1,2} E_T^{\rm{jet}} e^{\eta^{\rm{jet}}}}{2 \xpom E_p}\enspace ,
\end{eqnarray*}

where the sums run over the two highest $E_T$ jets.

In direct-photon processes, at LO in QCD, $x_{\gamma}$ is equal to one, 
whereas resolved-photon processes appear at $x_{\gamma} <1$. 
A direct-enriched region was defined by $x_{\gamma}^{\rm{obs}} \ge 0.75$ 
and a resolved-enriched region by $x_{\gamma}^{\rm{obs}} < 0.75$.

\section{Event selection}
\label{sec-evrecon}

A three-level trigger system was used to select events online 
\cite{lit-ZEUSdet,nim:a355:278,*Smith:1992b}.
Events with a large energy deposit in the calorimeter, neglecting the 
three inner rings of cells around the beampipe in the FCAL, were selected 
at the first-level trigger. 
Additional cuts were applied at the second-level trigger to reject 
beam-gas interactions and other non-$ep$ background events. At the 
third level,  the measured transverse energy, excluding  
the first inner ring of the FCAL, was required to be greater than 
$11\gev$. Jets were not preselected at any trigger level.

Well-reconstructed events were selected by applying the following 
quality cuts. The events were required to have at least three 
well-measured tracks of transverse momentum $p_T>0.2\,\gev$ 
originating from the same vertex.
The  longitudinal position of the vertex, $Z_{\rm{vtx}}$, had to be 
in the range $-35 < Z_{\rm{vtx}} < 30\,\rm{cm}$.  

Photoproduction events were selected as follows.
Events with a  scattered electron candidate having an inelasticity of 
$y_e \le 0.7$ were assumed to be DIS events and removed. In addition, 
$0.20 < y_{\rm{J\!B}} < 0.85$ was required. The cut on $y_e$ and the 
upper cut on $y_{\rm{J\!B}}$ reduced the remaining background from DIS events 
and also restricted the range of the virtuality of the exchanged photon 
to $Q^2 < 1 \gev^2$ with a median value of $10^{-3} \gev^2$.  
The lower cut on $y_{\rm{J\!B}}$
removed  proton-beam gas events which
deposit energy in the FCAL near the beam pipe.

Events with at least two jets were selected by requiring a transverse jet 
energy above $E_T^{\rm{jet}1(2)}>7.5\,(6.5) \gev$. 
Both jets were  required to be in the pseudorapidity range
$-1.5 < \eta^{\rm{jet1,2}} < 1.5$, measured in the laboratory frame.

Diffractive events were selected by requiring the presence of a LRG 
between the scattered proton and the rest of the hadronic final state. 
Since the proton was not measured, the requirement of a LRG was implemented 
by a cut on the total energy in the FPC, $E_{\rm{FPC}}<1.0\gev$, and 
by demanding  
$\eta_{\max} < 2.8$.      
Here $\eta_{\max}$ is defined as the pseudorapidity of the most forward 
EFO with an energy above $400\mev$ in the CAL. This selection ensures at 
least a two-unit rapidity gap in the hadronic system, suppressing background 
from non-diffractive and proton-dissociative processes.  
In addition, a cut  
$\xpom < 0.025$ was applied to enhance  the Pomeron-exchange 
contribution \cite{lit-golec}.

Finally, cosmic-ray events originating from muons that 
traversed the detector   
near the interaction point were removed. 
Details can be found elsewhere \cite{lit-SK,lit-renner}.
A total of 6990 events remained after all selection cuts.

\section{Monte Carlo simulations}
\label{subsec-theory}

  Monte Carlo (MC) simulations were used to 
determine acceptances and resolution effects at detector level  
and  to extract the hadronisation corrections for the NLO 
predictions, i.e. ratios of event yields at hadron level to those 
at parton level. 
 
The MC generator {\sc Rapgap}\cite{cpc:86:147} 
was used to simulate dijet processes in diffractive photoproduction
at the Born level. 
Electroweak radiative effects  were simulated by using 
{\sc Rapgap} in conjunction with {\sc Heracles 4.6} \cite{lit-heracles}.
 
The electron-proton interactions at small $Q^2$ were modelled with both 
direct and resolved photon processes (Fig.~\ref{fig-diagrams}).
Events were generated assuming that diffractive processes proceed via 
the emission of a particle-like Pomeron from the proton followed by the
interaction of the virtual photon with the Pomeron.  Although this
factorised approach has no justification in QCD, it gives a fair description 
of the data. 
The diffractive PDFs, as determined by the H1 Collaboration (H1 LO fit 2) 
\cite{zfp:c76:613} for the Pomeron contribution, were used. For resolved 
photon processes, the photon PDFs GRV-G-HO \cite{lit-GRV} 
were chosen.

In the simulation chain, the process of QCD radiation is followed by 
hadronisation. This was simulated by interfacing {\sc Rapgap} to a 
parton-shower model as implemented in {\sc Meps} 
\cite{zfp:c37:465} and to a hadronisation model based on string 
fragmentation \cite{cpc:39:347,cpc:43:367} as implemented in 
{\sc Jetset} \cite{cpc:46:43,*cpc:82:74}. 

The generated MC events were passed through the standard 
simulation of the ZEUS detector, based on the {\sc Geant} program 
\cite{tech:cern-dd-ee-84-1}, and a trigger-simulation package 
\cite{nim:a355:278,*Smith:1992b}. The simulated events were 
reconstructed and selected in the same way as the data.

Since the MC events generated with {\sc Rapgap} did not adequately 
describe the $\zpom$ distribution of the data, they were reweighted to the 
measured distribution separately for $\xgam\ge 0.75$ and $\xgam < 0.75$. 
The relative fractions of direct photon and resolved photon processes were 
determined from a fit  to the data. Resolved processes account
 for about one third  of the total event sample. 

\begin{figure}[p]
  \begin{center}
    \includegraphics[width=0.82\textwidth]{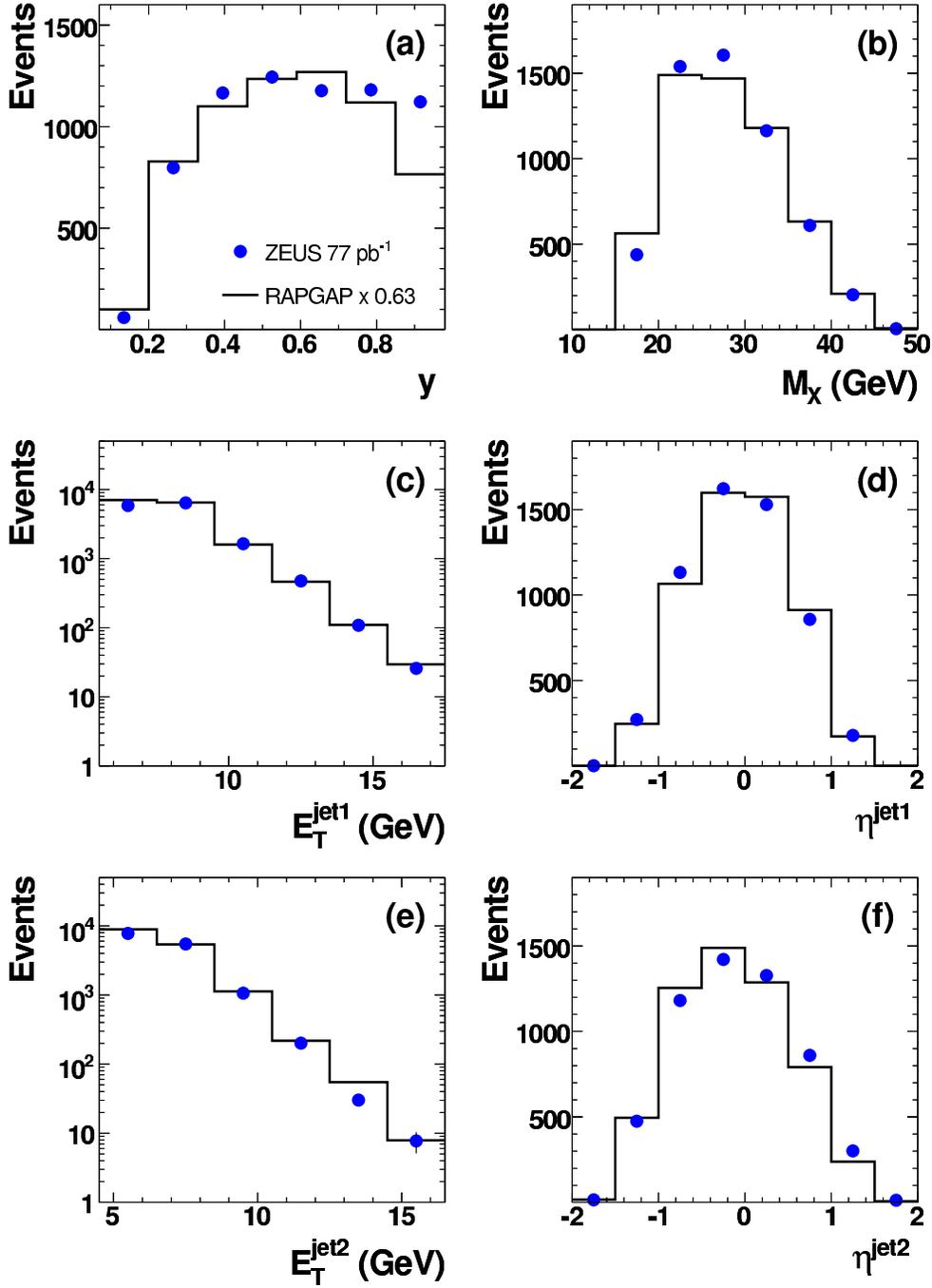}
  \end{center}
   \caption{Comparison of the data {\sl (dots)}\/ with the {\sc Rapgap MC} 
   {\sl (solid line)}\/ normalised to the data as a function of 
    {\sl (a)}\/ $y$, {\sl (b)}\/ $M_X$, {\sl (c)}\/ 
   $E_T^{\rm jet1}$, {\sl (d)}\/ $\eta^{\rm jet1}$,
   {\sl (e)}\/ $E_T^{\rm jet2}$ and {\sl (f)}\/ $\eta^{\rm jet2}$
   after all cuts but the one on the plotted variable.}
   \label{fig-control}
   \vfill
\end{figure}

Event distributions  are compared  with the reweighted {\sc Rapgap} 
MC distributions for the kinematic variables $y$, $M_X$, 
$E_T^{\rm{jet}1,2}$ and $\eta^{\rm{jet}1,2}$ in Fig.~\ref{fig-control}. 
The MC distributions were normalised to the data yielding a reasonable 
overall description of the data. 

The hadronisation corrections were calculated with the {\sc Rapgap} 
MC sample after reweighting its parton level $\zpom$ distribution  
to each of the NLO predictions described in Section~\ref{subsec-vsnloqcd}.
In addition, hadronisation corrections were also calculated with a 
MC sample generated with {\sc Pomwig} \cite{cpc:144:104}, 
a modification of the {\sc Herwig} MC program \cite{jhep:01:010} 
based on a cluster fragmentation model 
\cite{lit-webber1,lit-schmell}.
Since only direct photon interactions can be simulated with {\sc Pomwig}, 
the comparison to {\sc Rapgap} was restricted to the range $\xgam \ge 0.75$.
The bin-by-bin differences between the corrections obtained with the two 
programs  give an indication of the systematic 
uncertainties due to  the hadronisation corrections \cite{lit-renner}.
 
The MC generator {\sc Pythia} \cite{manual:cern-th-7112/93} was used to 
model the non-diffractive photoproduction of two jets. 
Events were generated using the CTEQ5L \cite{epj:c12:375} (GRV-G-HO) 
parametrisation of the proton (photon) PDFs and processed through the 
same simulation and selection chain as the data.

\section{Background}
\label{subsec-background}

Background from proton-dissociative events, with a low-mass 
proton-dissociative system 
escaping down the beam pipe, was estimated to be $(16\pm 4)\%$ \cite{lit-SK} 
by fitting to  the FPC energy distribution, 
without the $E_{FPC}$ cut, a mixture of 
{\sc Rapgap} and {\sc Epsoft} MC \cite{desy-95-069}. 
This value was also  
obtained from hard diffractive production of open charm \cite{np:b672:3}.
It was assumed that this estimate is independent of all kinematic variables 
studied here.  
The measured cross sections were scaled down accordingly.

Background  from non-diffractive dijet photoproduction, as estimated 
with the {\sc Pythia} MC, was found to be  less than $ 5\%$ throughout the
whole kinematic range, and was neglected.

\section{Systematic uncertainties}
\label{subsec-exp-uncertainties}

Systematic uncertainties on the measured cross sections 
were estimated as described below: 
\begin{itemize}
   \item the trigger efficiency was 
    estimated for both data and Monte Carlo events using an independent 
    trigger branch. The efficiency was above $ 98\%$ for the entire 
    kinematic range. The Monte Carlo simulation agrees with the data 
    within $\pm 1\%$ \cite{lit-renner} and the uncertainty was neglected;
   \item the transverse jet-energy scale was varied by $\pm 3\%$,
    the typical uncertainty in this $E_T^{\rm jet}$ range 
    \cite{lit-previous}.
    This variation  resulted  in  an uncertainty of less than 
    $\pm 5\%$;
   \item the FPC energy cut was varied by $\pm 0.5\gev$, resulting in an 
    uncertainty less than $\pm 1\%$ in most bins and not more than $\pm 2\%$; 
   \item changing the energy threshold of the EFOs, which is used to 
    calculate $\eta_{\rm max}$, by $\pm 100\mev$ led to an uncertainty 
    typically less than $\pm 1\%$ and not more than $\pm 2\%$ in any bin;
   \item the $\eta_{\rm max}$ values of data and Monte Carlo events were 
    shifted relative to each other by $\pm0.1$, 
    the typical $\eta$ resolution. This led to the largest 
    observed uncertainties which were typically below $\pm 6.5\%$ and up 
    to $\pm 14\%$ for low $\zpom$ and large $\xpom$ and $M_X$; 
   \item the lower $y_{\rm{J\!B}}$ cut was varied within its resolution 
    (0.04); the resulting uncertainties were typically less than $\pm 1\%$ 
    and not more than $\pm 3\%$. When the higher $y_{\rm{J\!B}}$ cut was 
    lowered, the measured cross sections changed typically by $<1\%$ and 
    not more than $\pm 4\%$;
   \item varying the  $\eta^{\rm{jet}}$ cuts within its resolution (0.1) 
    gave an 
    uncertainty which is mostly below $\pm 1\%$ and not more than $\pm 4\%$; 
   \item the $\xpom$ cut was varied within its resolution $(0.0025)$; 
    the resulting uncertainties were typically less than $\pm 2\%$, 
    increasing to $\pm 6\%$ in the highest $M_X$ and $\eta^{\rm{jet}1}$ bins. 
\end{itemize}

\begin{figure}[p]
  \begin{center}
    \includegraphics[width=0.65\textwidth]{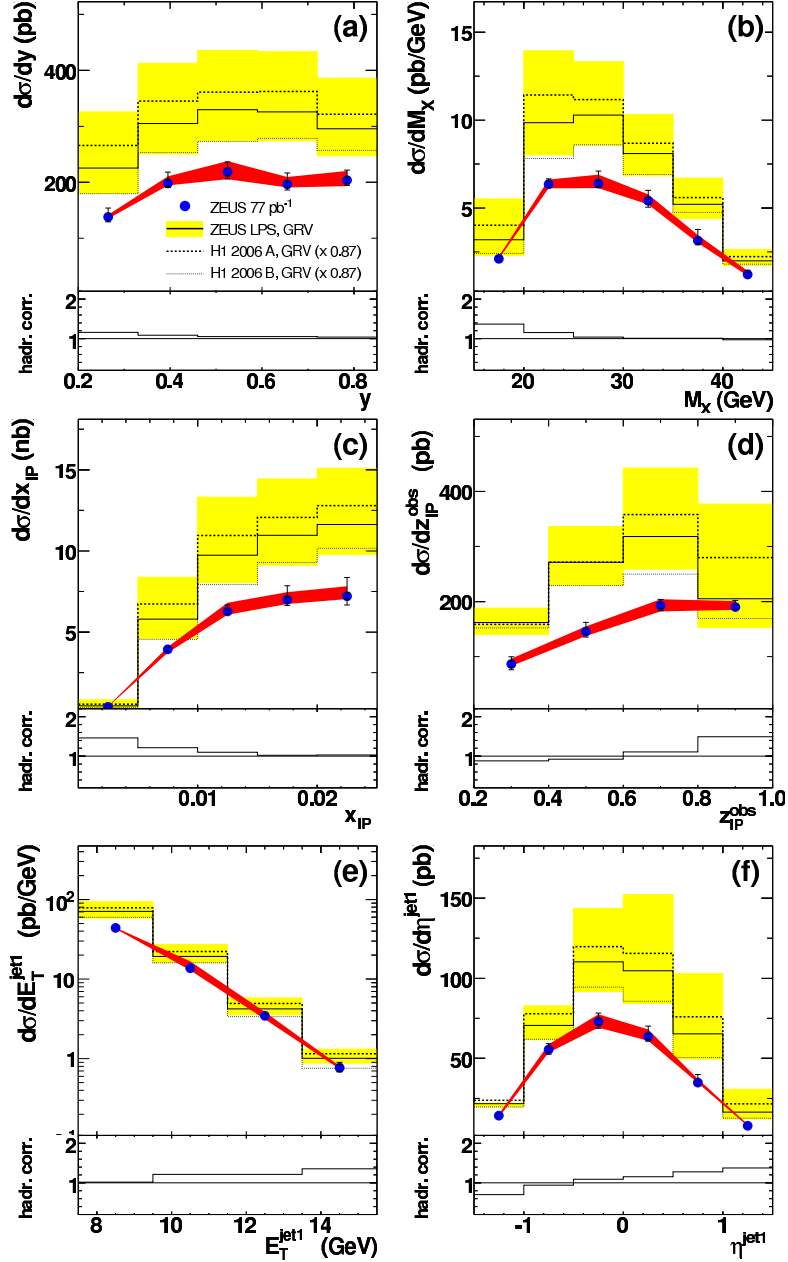}
  \end{center}
  \caption{Single-differential cross sections {\sl (dots)}\/ as a function
     of {\sl (a)}\/ $y$, {\sl (b)}\/ $M_X$, {\sl (c)}\/ 
     $\xpom$, {\sl (d)}\/ $\zpom$, 
     {\sl (e)}\/ $E_T^{\rm jet1}$ and {\sl (f)}\/ $\eta^{\rm jet1}$
     compared with NLO QCD predictions, corrected for 
     hadronisation, using the dPDFs
     from the
     ZEUS LPS fit {\sl (solid line),}\/ the H1 2006 fit A 
     {\sl (dashed line)}\/ and the H1 2006 fit B {\sl (dotted line)}\/
     and the GRV $\gamma$-PDFs.
     The inner error bars of the dots show the statistical
     uncertainty, the outer error bars show the statistical and  
     systematic uncertainties (see Section~\protect\ref{subsec-exp-uncertainties})
     added in quadrature. 
     The dark shaded band indicates the jet energy scale uncertainty. 
     The light shaded band shows the theoretical uncertainty due to the 
     variation of the scale when using the ZEUS LPS fit.
     Underneath each plot the hadronisation corrections applied to the 
     NLO prediction at parton level are shown. 
     }
    \label{fig-vsnlo_rest}
   \vfill
\end{figure}

\begin{figure}[p]
  \begin{center}
    \includegraphics[width=0.95\textwidth]{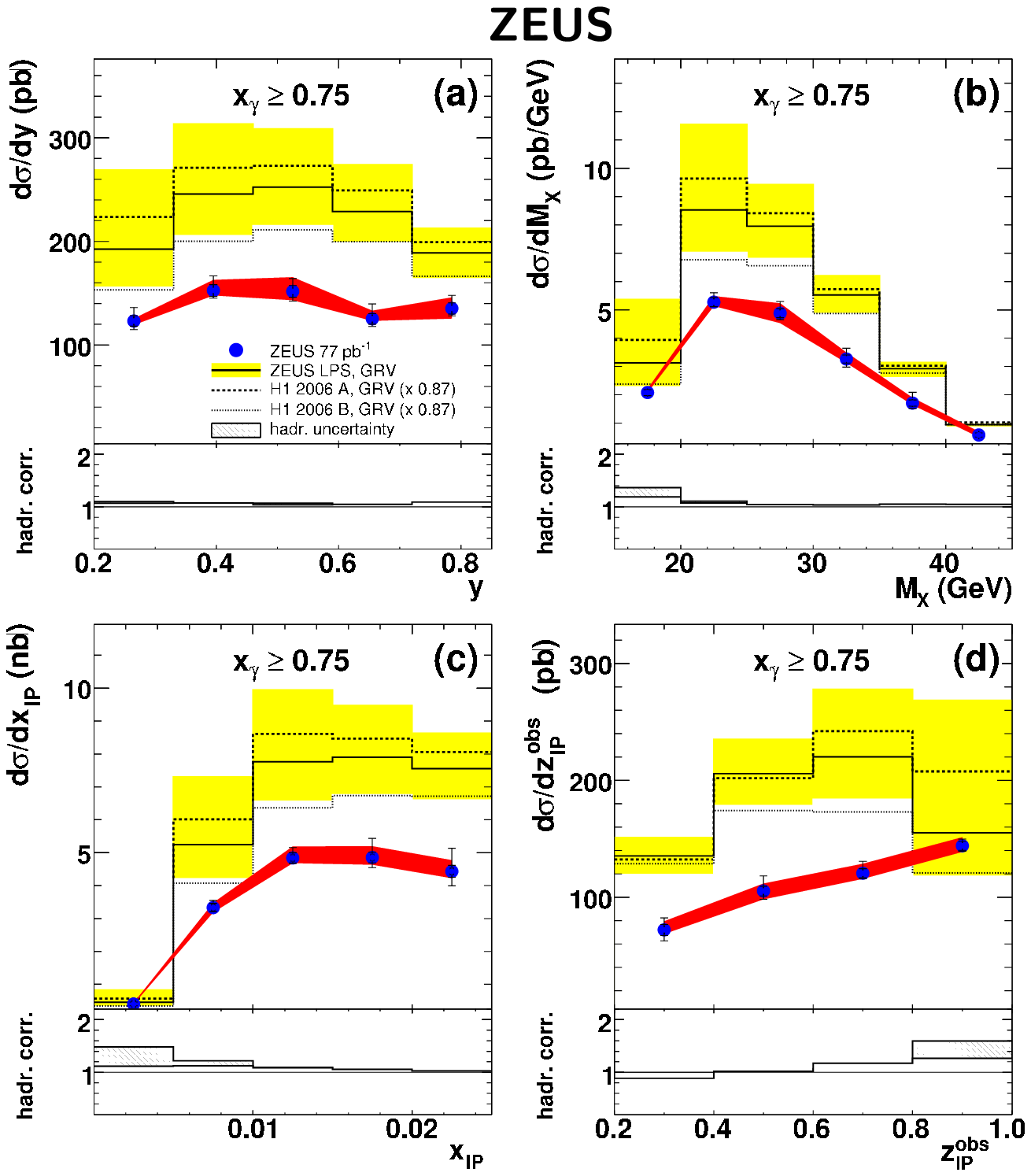}
  \end{center}
     \caption{Single-differential cross sections as a function of {\sl (a)}\/ 
     $y$, {\sl (b)}\/ $M_X$, {\sl (c)}\/~$\xpom$ 
     and {\sl (d)}\/ $\zpom$ 
     for direct-photon-enriched dijet photoproduction ($\xgam \ge 0.75$) 
     compared with NLO QCD predictions, corrected for hadronisation, using 
     the dPDFs from the ZEUS LPS fit (solid line), the H1 2006 fit A 
     (dashed line) and the H1 fit B (dotted line) and the GRV $\gamma$-PDFs.  
     Underneath each plot
     hadronisation corrections are shown which were obtained with {\sc Rapgap}
     (upper histogram) and {\sc Pomwig} (lower histogram), respectively. The 
      shaded bands 
     indicate the differences. The corrections from {\sc Rapgap} 
     were applied to obtain  
     the NLO predictions shown above.
     Further details are the same as in the caption of 
     Fig.~\ref{fig-vsnlo_rest}.}
      \label{fig-vslomc_larger0.75}
   \vfill
\end{figure}

\begin{figure}[p]
  \begin{center}
    \includegraphics[width=0.95\textwidth]{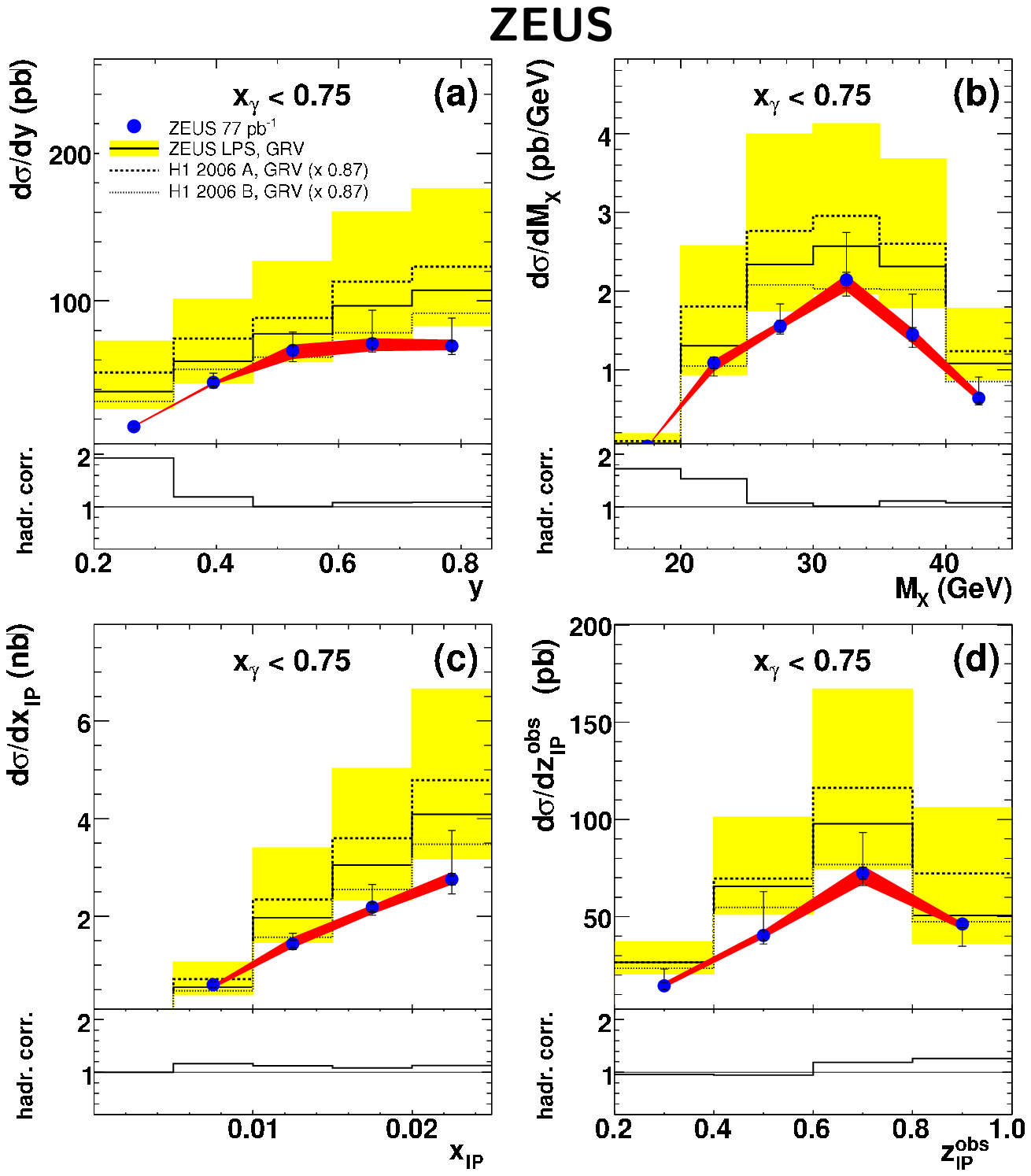}
  \end{center}
      \caption{Single-differential cross sections as a function of {\sl (a)}\/
     $y$, {\sl (b)}\/ $M_X$, {\sl (c)}\/~$\xpom$ 
     and {\sl (d)}\/ $\zpom$
      for resolved-photon-enriched dijet 
      photoproduction ($\xgam < 0.75$)  compared with 
      NLO QCD predictions, corrected for hadronisation, using the dPDFs from 
      the ZEUS LPS fit (solid line), the H1 2006 fit A (dashed line) and
      the H1 2006 fit B (dotted line) and the GRV $\gamma$-PDFs. Underneath 
      each plot the hadronisation corrections applied to the NLO predictions
      at parton level are shown.
      Further details are the same as in the caption of 
      Fig.~\ref{fig-vsnlo_rest}.}
      \label{fig-vslomc_lower0.75}
   \vfill
\end{figure}

The systematic uncertainties not associated with the jet-energy scale 
were added in quadrature to the statistical uncertainty and
are shown as error bars of the measured cross sections 
in Figs~\ref{fig-vsnlo_rest} to~\ref{fig-vslomc_lower0.75}. The uncertainty 
due to the energy scale 
is shown separately as a shaded band in each of the figures. 
Overall normalisation 
uncertainties of $\pm 2.2\%$ from the luminosity determination 
and of $\pm 4\%$ from 
subtraction of the dissociative background were not included.

\section{Results}
\label{sec-results}
\subsection{Cross sections}\label{subsec-vslomc}

Single-differential cross sections were measured in the kinematic region 
$Q^2 < 1\gev^2$, $0.20 < y < 0.85$, $\xpom < 0.025$, 
$E_T^{\rm{jet1(2)}} > 7.5\,(6.5)\gev$ and 
$-1.5< \eta^{\rm{jet1,2}} < 1.5$, 
and were determined as a function of $y$, $M_X$, $\xpom$, $\zpom$, 
$E_T^{\rm{jet}1}$, $\eta^{{\rm jet}1}$ and $\xgam$.  
The estimated contribution of proton-dissociative background of $16\%$ 
was subtracted in all bins.
 
\begin{figure}[p]
  \begin{center}
    \includegraphics[width=0.58\textwidth]{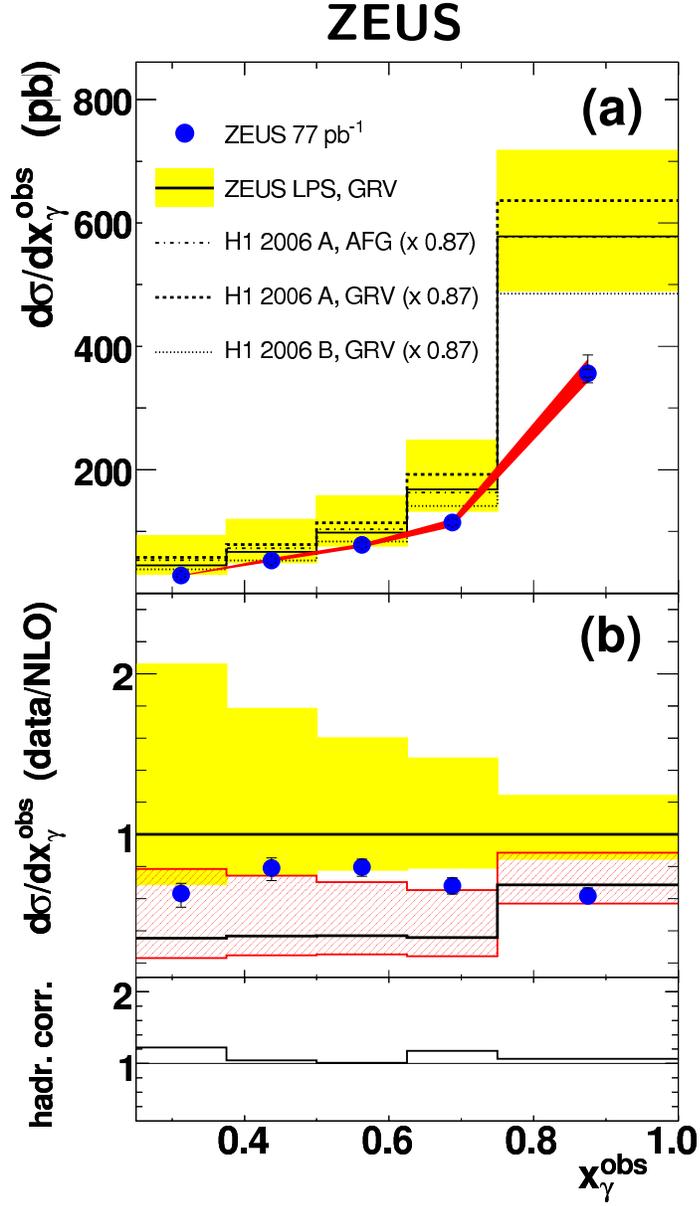}
  \end{center}
  \caption{{\sl (a)}\/ Single-differential cross section as a function of 
    $\xgam$ 
    compared with NLO QCD predictions, corrected for hadronisation, using 
    the dPDFs from the ZEUS LPS fit (solid line), the H1 2006 fit A 
    (dashed line)
    and the H1 2006 fit B (dotted line) and the GRV $\gamma$-dPDF.  
    The prediction with H1 2006 fit A is also shown using 
    the AFG parametrisation of the $\gamma-$PDFs {\sl (dashed-dotted line)}\/. 
    Other details are the same as in the caption of Fig.~\ref{fig-vsnlo_rest}.
    {\sl (b)} Ratio of  data and  NLO predictions 
    using the ZEUS LPS fit and GRV. The histogram  
    indicates the expectation with the predicted 
    resolved photon component scaled down by a factor of 0.34. 
    The shaded and hatched bands show the theoretical uncertainty.
    Underneath the hadronisation corrections applied to the 
    NLO prediction at parton level are shown.}
     \label{fig-xgamma}
     \vfill
\end{figure}

The cross sections are shown in Figs~\ref{fig-vsnlo_rest} 
and~\ref{fig-xgamma} and  listed 
in Tables~\ref{tab-fullxg}--\ref{tab-xgamma}.
The cross section dependence on $\xgam$, shown in Fig.~\ref{fig-xgamma}, 
indicates that direct-enriched ($\xgam \ge0.75$) processes 
dominate  diffractive dijet photoproduction in the kinematic range 
of this measurement. 
 
Single-differential cross sections were also determined separately 
for direct-photon enriched and resolved-photon enriched processes. 
They are shown in Figs~\ref{fig-vslomc_larger0.75} 
and~\ref{fig-vslomc_lower0.75}, 
respectively, and listed in Tables~\ref{tab-dir} and~\ref{tab-res}.  
The two sets of distributions differ in shape. Typically, resolved events are 
characterised by larger diffractive masses $M_X$; this  in turn  reflects    
into the observed $\xpom$ behaviour. 
Slight differences are observed in the $\zpom$ distributions 
with the most prominent feature being the rise of the direct-enriched 
component when 
$\zpom$ approaches one.
   
\subsection{Comparison to the NLO QCD calculations}
\label{subsec-vsnloqcd}

NLO predictions for diffractive photoproduction  of dijets were 
calculated at parton level with a program 
by  M.~Klasen and G.~Kramer \cite{lit-klasen}. 
The 
calculations were performed 
with a fixed-flavour number of $N_f=4$  
and $\Lambda_4 = 330\mev$, chosen to match the value of the running 
$\alpha_S$ in the region of four active flavours.
Three sets of  dPDFs were used: the  
ZEUS LPS fit, determined from an NLO QCD fit to inclusive diffraction and 
diffractive charm-production 
data \cite{epj:c38:43}, and the H1 2006 fits A and B, obtained from 
fits to inclusive diffraction data \cite{lit-D6}. The Regge-inspired 
parameters set for the NLO calculations were the 
same as used to obtain the dPDFs. The t-slope used in the Pomeron flux 
was 5$\gev^2$. For comparison 
with data, the NLO calculations obtained with the H1 dPDFs were scaled 
down by a factor\footnote{
The H1 measurements used to derive the H1 dPDFs include low-mass 
proton-dissociative processes which leads to an overestimate 
of the photon-diffractive cross section by a factor of 
 $(1.15^{+0.15}_{-0.08})$ as obtained from MC simulations~\cite{lit-D6}.}
of 0.87 \cite{lit-D6}. 
The contribution of subleading 
Regge trajectories as implemented in the H1 fits was included. For the 
resolved photon, the    
$\gamma$-PDF parametrisations GRV    
\cite{lit-GRV} and AFG04 \cite{lit-AFG} were used.     

The  NLO QCD predictions were obtained setting 
the renormalisation and factorisation scales  to 
 $\mu_R = \mu_F = \mu = E_T^{\rm jet1}$.  
The theoretical uncertainties  were estimated  
by varying the scales simultaneously between 
${(0.5 \cdot E_T^{\rm{jet}1})}$ and ${(2 \cdot E_T^{\rm{jet}1})}$
\cite{lit-klasen}. 
 Changing the number of active flavours to $N_f = 5$ in the NLO calculations 
leads to an increase of the expected cross section for $\xgam \ge 0.75$ by 
less than $10 \%$, and to a negligible effect elsewhere.
The uncertainties of the dPDFs and the Pomeron  flux, constraining 
directly the normalisation, were not included.
The predicted cross sections were transformed to the hadron level 
using the hadronisation corrections calculated with {\sc Rapgap} 
as described in Section~\ref{subsec-theory}. The uncertainties of 
the hadronisation corrections are not included in the error calculations 
for the cross sections. 
 
The data are compared with NLO QCD predictions at hadron level for the 
full $\xgam$ range in Figs~\ref{fig-vsnlo_rest} and~\ref{fig-xgamma}. 
The hadronisation corrections applied to the NLO predictions at parton 
level are shown in the lower part of each plot and the values are given  
in Tables~\ref{tab-fullxg}--\ref{tab-xgamma}. 
The asymmetric theoretical uncertainties, estimated as described, were 
determined  
 for the ZEUS LPS fit; 
those for the other NLO predictions are similar. The data are reasonably 
well described  in  shape. However, they lie systematically 
below  all the predictions. Most of the suppression  originates from 
the lower $E_T^{\rm {jet1}}$ region.
 
Figure~\ref{fig-xgamma}(b)  shows the ratio of the data and the NLO
predictions using the ZEUS LPS fit. The ratio is below one, consistent
with a  suppression factor of about 0.7 independent of $x_{\gamma}$.  
Also shown is the ratio expected   if the calculated 
 resolved-photon cross section is suppressed by a factor of $0.34$ 
\cite{lit-D14,*lit-D15}. No additional suppression factor for 
resolved-enriched data is observed.  
The suppression factor depends on the dPDFs and ranges between about 0.6 
(H1 2006 fit A) and about 0.9 (H1 2006 fit B). Within the large uncertainties 
of NLO calculations, the data are compatible with no suppression, as
expected in ~\cite{lit-bialas}.
 
Differential cross sections for the direct-enriched and resolved-enriched 
samples
are compared  with NLO predictions at hadron level
in Figs~\ref{fig-vslomc_larger0.75} and~\ref{fig-vslomc_lower0.75}.
Again the hadronisation corrections are shown in the lower part of each 
plot and the values are given in Tables~\ref{tab-dir} and~\ref{tab-res}. 
For direct-enriched data, the hadronisation corrections are shown for 
both {\sc Rapgap} and {\sc Pomwig}. The differences are taken as an 
estimate of 
the uncertainties as described in Section~\ref{subsec-theory}. The data lie 
systematically below the NLO calculations. Also, contrary to 
NLO expectations,   
the cross section as a function of $\zpom$ for the direct-enriched sample  
rises steadily with increasing $\zpom$.

Compared to NLO calculations obtained with the program of 
Frixione and Ridolfi\cite{lit-frix3},  
the H1 Collaboration observed a  suppression factor of about 0.5   
in both resolved-enriched and direct-enriched cross sections 
of diffractive dijet photoproduction \cite{lit-H1recent}. The
measurements of ZEUS and H1 cover different kinematic regions in $E_T$ and 
$\xpom$.\footnote{It was checked that
both programs for calculating  NLO predictions give consistent results.} 
In particular, the H1 measurements extend to lower  $E_T$ values than in the 
ZEUS analysis. In ZEUS, the largest discrepancy between the measured and 
predicted values of the cross section is observed at the lowest $E_T$ values   
suggesting that the conclusion on factorisation breaking depends on 
the probed scale.

\section{Conclusions}
\label{sec-conclusions}

Cross sections for
diffractive photoproduction of dijets were measured with the ZEUS detector 
at HERA using an integrated luminosity of $77.2 \pbi$. The measurements 
were performed in the kinematic region $Q^2 < 1\gev^2$, $0.20 < y < 0.85$ 
and $\xpom < 0.025$. The two jets with highest transverse energy were 
required to have $E_T^{\rm{jet1(2)}} > 7.5\,(6.5)\gev$ and 
$-1.5< \eta^{\rm{jet1,2}} < 1.5$. 
 
The measured differential cross sections  are compared to NLO QCD 
predictions based on available parameterisations of diffractive PDFs.
The comparisons were made for the full data sample as well as for the 
subsamples enriched with   
resolved photon  ($\xgam < 0.75$) and direct photon ($\xgam \ge 0.75$) 
processes. The NLO calculations tend to overestimate 
the measured cross sections of both the resolved-enriched and the 
direct-enriched 
data sample. However, within the large uncertainties of the NLO calculations 
the data are compatible with QCD factorisation.


\section{Acknowledgements}
\label{sec-ackn}

We are grateful to the DESY directorate for their strong support and 
encouragment. The effort of the HERA machine group is gratefully 
acknowledged. We thank the DESY computing and network services for their 
support. The design, construction and installation of the ZEUS detector 
has been made possible by the efforts of many people not listed as authors. 
It is a pleasure to thank M.~Klasen and G.~Kramer for handing over to us 
their program for calculating the NLO predictions and for carrying out
additional checks. 

\vfill\eject

\clearpage
{\raggedright
\providecommand{\etal}{et al.\xspace}
\providecommand{\coll}{Coll.\xspace}
\catcode`\@=11
\def\@bibitem#1{%
\ifmc@bstsupport
  \mc@iftail{#1}%
    {;\newline\ignorespaces}%
    {\ifmc@first\else.\fi\orig@bibitem{#1}}
  \mc@firstfalse
\else
  \mc@iftail{#1}%
    {\ignorespaces}%
    {\orig@bibitem{#1}}%
\fi}%
\catcode`\@=12

}

\begin{table}[h]
\begin{center}
\begin{tabular}{|c|cccc|c|}
\hline
$y$ bin & $d\sigma/dy$  & $\delta_{\rm{stat}}$ & $\delta_{\rm{syst}}$ & $\delta_{\rm{ES}}$ & $C_{\rm{had}}$ \\
$\,$ & $\,\rm{(pb)}\,$ & $\,\rm{(pb)}\,$ & $\,\rm{(pb)}\,$ & $\,\rm{(pb)}\,$ & \\
\hline
$0.20,\, 0.33$ & $137.9$ & $5.3$ & $ ^{+15.2}_{-7.0}$ & $ ^{+4.0}_{-2.9}$ & $1.16$ \\
$0.33,\, 0.46$ & $198.4$ & $6.3$ & $ ^{+18.9}_{-5.0}$ & $ ^{+12.1}_{-4.1}$ & $1.09$ \\
$0.46,\, 0.59$ & $218.3$ & $6.7$ & $ ^{+16.5}_{-10.4}$ & $ ^{+19.4}_{-12.8}$ & $1.05$ \\
$0.59,\, 0.72$ & $196.5$ & $6.2$ & $ ^{+19.1}_{-8.2}$ & $ ^{+13.0}_{-5.7}$ & $1.05$ \\
$0.72,\, 0.85$ & $203.6$ & $6.4$ & $ ^{+17.1}_{-6.6}$ & $ ^{+16.6}_{-10.4}$ & $1.04$ \\
\hline
$M_{X}$ bin & $d\sigma/dM_{X}$  & $\delta_{\rm{stat}}$ & $\delta_{\rm{syst}}$ & $\delta_{\rm{ES}}$ & $C_{\rm{had}}$ \\
$\,\rm{(GeV)}\,$ & $\,\rm{(pb/GeV)}\,$ & $\,\rm{(pb/GeV)}\,$ & $\,\rm{(pb/GeV)}\,$ & $\,\rm{(pb/GeV)}\,$ & \\
\hline
$15.0,\, 20.0$ & $2.11$ & $0.11$ & $ ^{+0.11}_{>-0.01}$ & $ ^{+0.13}_{>-0.01}$ & $1.37$ \\
$20.0,\, 25.0$ & $6.35$ & $0.17$ & $ ^{+0.24}_{-0.13}$ & $ ^{+0.26}_{-0.23}$ & $1.16$ \\
$25.0,\, 30.0$ & $6.39$ & $0.17$ & $ ^{+0.68}_{-0.15}$ & $ ^{+0.51}_{-0.26}$ & $1.04$ \\
$30.0,\, 35.0$ & $5.41$ & $0.17$ & $ ^{+0.57}_{-0.32}$ & $ ^{+0.36}_{-0.21}$ & $1.01$ \\
$35.0,\, 40.0$ & $3.14$ & $0.14$ & $ ^{+0.62}_{-0.19}$ & $ ^{+0.32}_{-0.13}$ & $1.01$ \\
$40.0,\, 45.0$ & $1.21$ & $0.09$ & $ ^{+0.23}_{-0.08}$ & $ ^{+0.21}_{-0.07}$ & $0.97$ \\
\hline
$\xpom$ bin & $d\sigma/d\xpom$  & $\delta_{\rm{stat}}$ & $\delta_{\rm{syst}}$ & $\delta_{\rm{ES}}$ & $C_{\rm{had}}$ \\
$\,$ & $\,\rm{(nb)}\,$ & $\,\rm{(nb)}\,$ & $\,\rm{(nb)}\,$ & $\,\rm{(nb)}\,$ & \\
\hline
$0.000,\, 0.005$ & $0.40$ & $0.06$ & $ ^{+0.01}_{-0.04}$ & $ ^{+0.04}_{>-0.01}$ & $1.46$ \\
$0.005,\, 0.010$ & $3.94$ & $0.14$ & $ ^{+0.18}_{-0.10}$ & $ ^{+0.14}_{-0.21}$ & $1.21$ \\
$0.010,\, 0.015$ & $6.28$ & $0.17$ & $ ^{+0.36}_{-0.10}$ & $ ^{+0.53}_{-0.20}$ & $1.10$ \\
$0.015,\, 0.020$ & $7.00$ & $0.19$ & $ ^{+0.84}_{-0.31}$ & $ ^{+0.47}_{-0.27}$ & $1.02$ \\
$0.020,\, 0.025$ & $7.21$ & $0.21$ & $ ^{+1.13}_{-0.50}$ & $ ^{+0.62}_{-0.18}$ & $1.03$ \\
\hline
$\zpom$ bin & $d\sigma/d\zpom$  & $\delta_{\rm{stat}}$ & $\delta_{\rm{syst}}$ & $\delta_{\rm{ES}}$ & $C_{\rm{had}}$ \\
$\,$ & $\,\rm{(pb)}\,$ & $\,\rm{(pb)}\,$ & $\,\rm{(pb)}\,$ & $\,\rm{(pb)}\,$ & \\
\hline
$0.2,\, 0.4$ & $86.4$ & $5.0$ & $ ^{+12.4}_{-9.0}$ & $ ^{+9.9}_{-3.6}$ & $0.88$ \\
$0.4,\, 0.6$ & $145.7$ & $4.9$ & $ ^{+15.7}_{-8.9}$ & $ ^{+9.3}_{-8.2}$ & $0.92$ \\
$0.6,\, 0.8$ & $192.9$ & $4.9$ & $ ^{+9.6}_{-8.0}$ & $ ^{+11.3}_{-10.7}$ & $1.11$ \\
$0.8,\, 1.0$ & $190.2$ & $4.2$ & $ ^{+11.2}_{-0.2}$ & $ ^{+10.8}_{-5.2}$ & $1.49$ \\
\hline
\end{tabular}
\caption{{\small Differential cross sections for the diffractive photoproduction 
of dijets as a function of $y$, $M_X$, $\xpom$ and $\zpom$ listed with  
statistical ($\delta_{\rm{stat}}$) and systematic ($\delta_{\rm{syst}}$)  
uncertainties  and  energy scale ($\delta_{\rm{ES}}$) uncertainties; the last 
column shows the hadronisation corrections 
($C_{\rm{had}}$) applied to the NLO QCD predictions.}}
\label{tab-fullxg}
\end{center}
\end{table}

\begin{table}[h]
\begin{center}
\begin{tabular}{|c|cccc|c|}
\hline
$E_{T}^{\rm{jet1}}$ bin & $d\sigma/dE_{T}^{\rm{jet1}}$  & $\delta_{\rm{stat}}$ & $\delta_{\rm{syst}}$ & $\delta_{\rm{ES}}$ & $C_{\rm{had}}$ \\
$\,\rm{(GeV)}\,$ & $\,\rm{(pb/GeV)}\,$ & $\,\rm{(pb/GeV)}\,$ & $\,\rm{(pb/GeV)}\,$ & $\,\rm{(pb/GeV)}\,$ & \\
\hline
$7.5,\, 9.5$ & $44.0$ & $0.8$ & $ ^{+3.9}_{-1.6}$ & $ ^{+0.4}_{-1.0}$ & $1.02$ \\
$9.5,\, 11.5$ & $13.7$ & $0.4$ & $ ^{+1.3}_{-0.5}$ & $ ^{+3.1}_{-0.7}$ & $1.22$ \\
$11.5,\, 13.5$ & $3.5$ & $0.2$ & $ ^{+0.3}_{-0.2}$ & $ ^{+0.4}_{-0.3}$ & $1.22$ \\
$13.5,\, 15.5$ & $0.8$ & $0.1$ & $ ^{+0.1}_{-0.0}$ & $ ^{+0.1}_{-0.1}$ & $1.35$ \\
\hline
$\eta^{\rm{jet1}}$ bin & $d\sigma/d\eta^{\rm{jet1}}$  & $\delta_{\rm{stat}}$ & $\delta_{\rm{syst}}$ & $\delta_{\rm{ES}}$ & $C_{\rm{had}}$ \\
$\,$ & $\,\rm{(pb)}\,$ & $\,\rm{(pb)}\,$ & $\,\rm{(pb)}\,$ & $\,\rm{(pb)}$  & \\
\hline
$-1.5,\, -1.0$ & $14.2$ & $0.9$ & $ ^{+1.5}_{-0.6}$ & $ ^{+1.0}_{-1.0}$ & $0.70$ \\
$-1.0,\, -0.5$ & $55.6$ & $1.8$ & $ ^{+3.1}_{-2.4}$ & $ ^{+3.2}_{-1.9}$ & $0.94$ \\
$-0.5,\, 0.0$ & $72.9$ & $2.0$ & $ ^{+5.0}_{-3.6}$ & $ ^{+4.7}_{-3.7}$ & $1.09$ \\
$0.0,\, 0.5$ & $63.6$ & $1.8$ & $ ^{+6.3}_{-2.2}$ & $ ^{+4.5}_{-2.6}$ & $1.16$ \\
$0.5,\, 1.0$ & $34.9$ & $1.3$ & $ ^{+4.9}_{-0.8}$ & $ ^{+2.3}_{-0.2}$ & $1.28$ \\
$1.0,\, 1.5$ & $8.0$ & $0.6$ & $ ^{+1.1}_{-0.3}$ & $ ^{+0.9}_{-0.3}$ & $1.38$ \\
\hline
\end{tabular}
\caption{{\small Differential cross sections for the 
diffractive photoproduction 
of dijets as a function of $E_T^{\rm{jet1}}$ and $\eta^{\rm{jet1}}$  
listed with 
statistical ($\delta_{\rm{stat}}$) and systematic ($\delta_{\rm{syst}}$) 
uncertainties and energy scale ($\delta_{\rm{ES}}$) uncertainties ; 
the last column shows the hadronisation corrections 
($C_{\rm{had}}$) applied to the NLO QCD predictions.}}
\label{tab-fullxg2}
\end{center}
\end{table}

\begin{table}[h]
\begin{center}
\begin{tabular}{|c|cccc|c|}
\hline
$\xgam$ bin & $d\sigma/d\xgam$  & $\delta_{\rm{stat}}$ & $\delta_{\rm{syst}}$ & $\delta_{\rm{ES}}$ & $C_{\rm{had}}$ \\
$\,$ & $\,\rm{(pb)}\,$ & $\,\rm{(pb)}\,$  & $\,\rm{(pb)}\,$  & $\,\rm{(pb)}\,$  & \\
\hline
$0.250,\, 0.375$ & $28.5$ & $2.4$ & $ ^{+1.4}_{-3.0}$ & $ ^{+0.5}_{-1.1}$ & $1.23$ \\
$0.375,\, 0.500$ & $52.7$ & $3.2$ & $ ^{+2.7}_{-4.1}$ & $ ^{+4.2}_{-1.9}$ & $1.04$ \\
$0.500,\, 0.625$ & $78.1$ & $3.7$ & $ ^{+3.2}_{-4.2}$ & $ ^{+1.9}_{-4.5}$ & $1.01$ \\
$0.625,\, 0.750$ & $114.3$ & $4.5$ & $ ^{+6.8}_{-7.4}$ & $ ^{+5.9}_{-7.3}$ & $1.18$ \\
$0.750,\, 1.000$ & $356.5$ & $6.2$ & $ ^{+29.1}_{-14.1}$ & $ ^{+23.3}_{-14.4}$ & $1.07$ \\
\hline
\end{tabular}
\caption[Cross sections and stat. $\pm$ syst. uncertainty for $\xgam$.]
{Differential cross sections  for the diffractive photoproduction of dijets  
as a function of $\xgam$ 
listed with statistical ($\delta_{\rm{stat}}$) and systematic 
($\delta_{\rm{syst}}$) uncertainties and energy scale ($\delta_{\rm{ES}}$) 
uncertainties;
 the last column shows the  hadronisation corrections applied to the NLO QCD
 predictions.}
\label{tab-xgamma}
\end{center}
\end{table}

\clearpage

\begin{table}[h]
\begin{center}
\begin{tabular}{|c|cccc|c|}
\hline
$y$ bin & $d\sigma/dy$  & $\delta_{\rm{stat}}$ & $\delta_{\rm{syst}}$ & $\delta_{\rm{ES}}$ & $C_{\rm{had}}$ \\
$\,$ & $\,\rm{(pb)}\,$ & $\,\rm{(pb)}\,$ & $\,\rm{(pb)}\,$  & $\,\rm{(pb)}\,$ 
& \\
\hline
$0.20,\, 0.33$ & $123.2$ & $5.1$ & $ ^{+12.0}_{-6.7}$ & $ ^{+2.3}_{-3.8}$ & $1.10$ \\
$0.33,\, 0.46$ & $152.6$ & $5.7$ & $ ^{+13.0}_{-4.5}$ & $ ^{+10.2}_{-4.8}$ & $1.07$ \\
$0.46,\, 0.59$ & $151.8$ & $5.7$ & $ ^{+11.0}_{-7.3}$ & $ ^{+13.6}_{-8.5}$ & $1.07$ \\
$0.59,\, 0.72$ & $125.2$ & $5.0$ & $ ^{+13.4}_{-5.3}$ & $ ^{+7.9}_{-1.8}$ & $1.05$ \\
$0.72,\, 0.85$ & $135.3$ & $5.4$ & $ ^{+11.3}_{-4.9}$ & $ ^{+10.6}_{-9.8}$ & $1.09$ \\
\hline
$M_{X}$ bin & $d\sigma/dM_{X}$  & $\delta_{\rm{stat}}$ & $\delta_{\rm{syst}}$ & $\delta_{\rm{ES}}$ & $C_{\rm{had}}$ \\
$\,\rm{(GeV)}\,$ & $\,\rm{(pb/GeV)}\,$ & $\,\rm{(pb/GeV)}\,$  
& $\,\rm{(pb/GeV)}\,$ & $\,\rm{(pb/GeV)}\,$ & \\
\hline
$15.0,\, 20.0$ & $2.08$ & $0.11$ & $ ^{+0.10}_{-0.01}$ & $ ^{+0.13}_{>-0.01}$ & $1.37$ \\
$20.0,\, 25.0$ & $5.27$ & $0.16$ & $ ^{+0.30}_{-0.08}$ & $ ^{+0.22}_{-0.13}$ & $1.11$ \\
$25.0,\, 30.0$ & $4.88$ & $0.16$ & $ ^{+0.38}_{-0.17}$ & $ ^{+0.35}_{-0.35}$ & $1.05$ \\
$30.0,\, 35.0$ & $3.26$ & $0.14$ & $ ^{+0.36}_{-0.25}$ & $ ^{+0.23}_{-0.14}$ & $1.03$ \\
$35.0,\, 40.0$ & $1.70$ & $0.11$ & $ ^{+0.36}_{-0.16}$ & $ ^{+0.19}_{-0.06}$ & $1.05$ \\
$40.0,\, 45.0$ & $0.58$ & $0.07$ & $ ^{+0.12}_{-0.07}$ & $ ^{+0.12}_{-0.07}$ & $1.05$ \\
\hline
$\xpom$ bin & $d\sigma/d\xpom$  & $\delta_{\rm{stat}}$ & $\delta_{\rm{syst}}$ & $\delta_{\rm{ES}}$ & $C_{\rm{had}}$ \\
$\,$ & $\,\rm{(nb)}\,$ & & & & \\
\hline
$0.000,\, 0.005$ & $0.40$ & $0.06$ & $ ^{+0.01}_{-0.05}$ & $ ^{+0.03}_{>-0.01}$ & $1.48$ \\
$0.005,\, 0.010$ & $3.33$ & $0.12$ & $ ^{+0.18}_{-0.05}$ & $ ^{+0.14}_{-0.15}$ & $1.22$ \\
$0.010,\, 0.015$ & $4.84$ & $0.15$ & $ ^{+0.27}_{-0.09}$ & $ ^{+0.34}_{-0.16}$ & $1.10$ \\
$0.015,\, 0.020$ & $4.86$ & $0.16$ & $ ^{+0.56}_{-0.26}$ & $ ^{+0.34}_{-0.22}$ & $1.06$ \\
$0.020,\, 0.025$ & $4.43$ & $0.18$ & $ ^{+0.68}_{-0.40}$ & $ ^{+0.34}_{-0.21}$ & $1.03$ \\
\hline
$\zpom$ bin & $d\sigma/d\zpom$  & $\delta_{\rm{stat}}$ & $\delta_{\rm{syst}}$ & $\delta_{\rm{ES}}$ & $C_{\rm{had}}$ \\
$\,$ & $\,\rm{(pb)}\,$ & $\,\rm{(pb/GeV)}\,$ & $\,\rm{(pb/GeV)}\,$ & $\,\rm{(pb/GeV})\,$ & \\
\hline
$0.2,\, 0.4$ & $72.0$ & $4.7$ & $ ^{+9.1}_{-8.1}$ & $ ^{+7.4}_{-3.3}$ & $0.89$ \\
$0.4,\, 0.6$ & $105.3$ & $4.2$ & $ ^{+12.3}_{-5.6}$ & $ ^{+6.5}_{-7.1}$ & $1.02$ \\
$0.6,\, 0.8$ & $120.6$ & $3.8$ & $ ^{+9.5}_{-3.1}$ & $ ^{+8.0}_{-4.8}$ & $1.17$ \\
$0.8,\, 1.0$ & $144.0$ & $3.8$ & $ ^{+4.4}_{-2.4}$ & $ ^{+7.3}_{-5.2}$ & $1.59$ \\
\hline
\end{tabular}
\caption[Cross sections with  stat. $\pm$ syst.  for $\xgam \ge 0.75$.]
{Differential cross sections for the diffractive  photoproduction of dijets 
for $\xgam \ge 0.75$ listed with 
 statistical ($\delta_{\rm{stat}}$) and systematic ($\delta_{\rm{syst}}$)  
uncertainties and energy scale ($\delta_{\rm{ES}}$) uncertainties; 
the last column shows the 
 hadronisation corrections applied to the NLO QCD predictions.
 }
\label{tab-dir}
\end{center}
\end{table}

\begin{table}[h]
\begin{center}
\begin{tabular}{|c|cccc|c|}
\hline
$y$ bin & $d\sigma/dy$  & $\delta_{\rm{stat}}$ & $\delta_{\rm{syst}}$ & $\delta_{\rm{ES}}$ & $C_{\rm{had}}$ \\
$\,$ & $\,\rm{(pb)}\,$ & $\,\rm{(pb)}\,$ & $\,\rm{(pb)}\,$  & $\,\rm{(pb)}\,$ & \\
\hline
$0.20,\, 0.33$ & $14.7$ & $1.5$ & $ ^{+2.5}_{-1.0}$ & $ ^{+0.7}_{-0.1}$ & $1.93$ \\
$0.33,\, 0.46$ & $44.9$ & $2.7$ & $ ^{+5.6}_{-2.7}$ & $ ^{+0.1}_{-1.7}$ & $1.19$ \\
$0.46,\, 0.59$ & $66.5$ & $3.5$ & $ ^{+11.9}_{-6.9}$ & $ ^{+3.9}_{-5.9}$ & $1.01$ \\
$0.59,\, 0.72$ & $71.0$ & $3.6$ & $ ^{+22.4}_{-4.4}$ & $ ^{+3.6}_{-5.0}$ & $1.08$ \\
$0.72,\, 0.85$ & $69.5$ & $3.5$ & $ ^{+18.6}_{-4.5}$ & $ ^{+4.3}_{-2.8}$ & $1.09$ \\
\hline
$M_{X}$ bin & $d\sigma/dM_{X}$  & $\delta_{\rm{stat}}$ & $\delta_{\rm{syst}}$ & $\delta_{\rm{ES}}$ & $C_{\rm{had}}$ \\
$\,\rm{(GeV)}\,$ & $\,\rm{(pb/GeV)}\,$ & $\,\rm{(pb/GeV)}\,$ & $\,\rm{(pb/GeV)}\,$ & $\,\rm{(pb/GeV)}\,$ & \\
\hline
$15.0,\, 20.0$ & $0.03$ & $0.01$ & $ ^{+0.01}_{-0.01}$ & $ ^{<+0.01}_{>-0.01}$ & $1.72$ \\
$20.0,\, 25.0$ & $1.09$ & $0.07$ & $ ^{<+0.01}_{-0.15}$ & $ ^{+0.04}_{-0.11}$ & $1.54$ \\
$25.0,\, 30.0$ & $1.55$ & $0.07$ & $ ^{+0.27}_{-0.07}$ & $ ^{+0.06}_{-0.04}$ & $1.07$ \\
$30.0,\, 35.0$ & $2.14$ & $0.10$ & $ ^{+0.59}_{-0.18}$ & $ ^{+0.07}_{-0.14}$ & $1.01$ \\
$35.0,\, 40.0$ & $1.45$ & $0.09$ & $ ^{+0.50}_{-0.14}$ & $ ^{+0.07}_{-0.12}$ & $1.11$ \\
$40.0,\, 45.0$ & $0.64$ & $0.06$ & $ ^{+0.26}_{-0.06}$ & $ ^{+0.07}_{-0.04}$ & $1.08$ \\
\hline
$\xpom$ bin & $d\sigma/d\xpom$  & $\delta_{\rm{stat}}$ & $\delta_{\rm{syst}}$ & $\delta_{\rm{ES}}$ & $C_{\rm{had}}$ \\
$\,$ & $\,\rm{(nb)}\,$ & $\,\rm{(nb)}\,$ & $\,\rm{(nb)}\,$ & $\,\rm{nb)}\,$ & \\
\hline
$0.005,\, 0.010$ & $0.60$ & $0.05$ & $ ^{<+0.01}_{-0.07}$ & $ ^{<+0.01}_{-0.07}$ & $1.16$ \\
$0.010,\, 0.015$ & $1.44$ & $0.08$ & $ ^{+0.20}_{-0.09}$ & $ ^{+0.13}_{-0.08}$ & $1.13$ \\
$0.015,\, 0.020$ & $2.19$ & $0.10$ & $ ^{+0.46}_{-0.13}$ & $ ^{+0.04}_{-0.14}$ & $1.08$ \\
$0.020,\, 0.025$ & $2.76$ & $0.12$ & $ ^{+0.99}_{-0.28}$ & $ ^{+0.15}_{-0.11}$ & $1.13$ \\
\hline
$\zpom$ bin & $d\sigma/d\zpom$  & $\delta_{\rm{stat}}$ & $\delta_{\rm{syst}}$ & $\delta_{\rm{ES}}$ & $C_{\rm{had}}$ \\
$\,$ & $\,\rm{(pb)}\,$ & $\,\rm{(pb)}\,$ & $\,\rm{(pb)}\,$ & $\,\rm{(pb)}\,$ & \\
\hline
$0.2,\, 0.4$ & $14.4$ & $1.8$ & $ ^{+8.5}_{-1.5}$ & $ ^{+1.8}_{-0.9}$ & $0.96$ \\
$0.4,\, 0.6$ & $40.4$ & $2.6$ & $ ^{+22.3}_{-3.7}$ & $ ^{+2.5}_{-1.4}$ & $0.95$ \\
$0.6,\, 0.8$ & $72.3$ & $3.1$ & $ ^{+20.7}_{-5.6}$ & $ ^{+3.8}_{-5.5}$ & $1.19$ \\
$0.8,\, 1.0$ & $46.2$ & $1.8$ & $ ^{+0.8}_{-11.3}$ & $ ^{+0.8}_{-2.7}$ & $1.26$ \\
\hline
\end{tabular}
\caption[Cross sections and stat. $\pm$ syst. uncertainty for $\xgam < 0.75$.]
{Differential cross sections for the  diffractive photoproduction of dijets  
for $\xgam < 0.75$ listed 
with  statistical ($\delta_{\rm{stat}}$) and systematic ($\delta_{\rm{syst}}$) 
uncertainties and energy scale ($\delta_{\rm{ES}}$) uncertainties; 
the last column shows
the  hadronisation corrections applied to the NLO QCD predictions.}
\label{tab-res}
\end{center}
\end{table}

%
%

\newpage
%
%

\vfill\eject

%
%

\begin{mcbibliography}{10}

\bibitem{lit-regge}
P.D.B.~Collins, {\it An Introduction to Regge Theory and High Energy Physics},
  Cambridge University Press (1977)\relax
\relax
\bibitem{lit-D1b}
A.C.~Irving and R.P.~Worden,
\newblock Phys.~Rept.{} {\bf 34},~117~(1977)\relax
\relax
\bibitem{lit-D1c}
A.B.~Kaidalov, {\it Regge Poles in {QCD},} in M.~Shifman (ed.): ``At the
  Frontier of Particle Physics", Vol. 1, p. 603, World Scientific (2002)\relax
\relax
\bibitem{pl:b323:201}
L.~Trentadue and G.~Veneziano,
\newblock Phys.\ Lett.{} {\bf B~323},~201~(1994)\relax
\relax
\bibitem{lit-D3}
A.~Berera and D.E.~Soper,
\newblock Phys.~Rev.{} {\bf D~53},~6162~(1996)\relax
\relax
\bibitem{lit-QCDfact}
J.C.~Collins, Phys. Rev. {\bf D 57}, 3051 (1998) ; Erratum ibid. {\bf D 61},
  019902 (2000)\relax
\relax
\bibitem{epj:c38:43}
ZEUS \coll, S.~Chekanov \etal,
\newblock Eur.\ Phys.\ J.{} {\bf C~38},~43~(2004)\relax
\relax
\bibitem{lit-D6}
H1~Collaboration, A.~Aktas \etal,
\newblock Eur.~Phys.~J.{} {\bf C~48},~749~(2006)\relax
\relax
\bibitem{lit-groys}
M.~Groys, A.~Levy and A.~Proskuryakov, Proc. of the workshop ``HERA and the
  LHC", H. Jung and A. De Roeck (eds.), CERN-2005-014, DESY-PROC-2005-001, p.
  499 (2005), also preprint hep-ph/0601012\relax
\relax
\bibitem{lit-D9}
A.D.~Martin, M.G.~Ryskin and G.~Watt,
\newblock Phys.~Lett.{} {\bf B~644},~131~(2007)\relax
\relax
\bibitem{lit-Dstar}
ZEUS Collaboration, S.~ Chekanov \etal,
\newblock Eur.~Phys.~J.{} {\bf C~51},~301~(2007)\relax
\relax
\bibitem{lit-H1recent}
H1~Collaboration, A.~Aktas \etal, Eur. Phys. J. {\bf D 51},~549 (2007)\relax
\relax
\bibitem{lit-ddis2}
H1 Collaboration, A.Aktas \etal, 
  {preprint} hep-ex/0610076, also {preprint} DESY-06-164\relax
\relax
\bibitem{lit-ddis1}
ZEUS~Collaboration, S. Chekanov \etal, submitted to Eur. Phys. J. {\bf C}
  (2007), also {preprint} DESY-07-126\relax
\relax
\bibitem{lit-LHC3}
D.~Graudenz and G.~Veneziano,
\newblock Phys.~Lett.{} {\bf B~365},~302~(1996)\relax
\relax
\bibitem{lit-LHC1}
L.~Alvero \etal,
\newblock Phys.~Rev.{} {\bf D~59},~0704022~(1999)\relax
\relax
\bibitem{lit-LHC4}
V.A.~Khoze, A.D.~Martin and M.G.~Ryskin,
\newblock Eur.~Phys.~J{} {\bf C~23},~311~(2002)\relax
\relax
\bibitem{lit-LHC2}
C.~Royon,
\newblock Acta~Phys.~Polon.{} {\bf B~37},~3571~(2006)\relax
\relax
\bibitem{lit-tevatron}
E. Gotsman \etal, {preprint} hep-ph/0511060, also Proc. of the ~workshop "HERA
  and the LHC", H. Jung and A. de Roeck (eds.), CERN-2005-014,
  DESY-PROC-2005-001, p. 221 (2005)\relax
\relax
\bibitem{lit-QCDfact2}
P.V.~Landshoff and J.C.~Polkinghorn, Nucl.~Phys. {\bf B~33}, 221 (1971) and
  {\bf B~36}, 642 (1972)\relax
\relax
\bibitem{lit-QCDfact3}
J.C.~Collins \etal,
\newblock Phys.~Lett.{} {\bf B~307},~161~(1993)\relax
\relax
\bibitem{prl:84:5043}
CDF \coll, T.~Affolder \etal,
\newblock Phys.\ Rev.\ Lett.{} {\bf 84},~5043~(2000)\relax
\relax
\bibitem{lit-D14}
A.B.~Kaidalov \etal,
\newblock Eur.~Phys.~J.{} {\bf C~21},~521~(2001)\relax
\relax
\bibitem{lit-D15}
A.B.~Kaidalov \etal,
\newblock Phys.~Lett.{} {\bf B~567},~61~(2003)\relax
\relax
\bibitem{lit-bialas}
A.~Bialas, prepared for 35th Int. Symposium on Multiparticle Dynamics (ISMD
  05), Kromeriz, Czech Republic, 9-15 Aug 2005; published in AIP Conf. Proc.
  {\bf 828}, 359 (2006); also in Kromeriz 2005, Multiparticle Dynamics,
  359\relax
\relax
\bibitem{lit-previous}
ZEUS~Collaboration, S.~Chekanov \etal, Phys. Lett. {\bf B~547}, 164
  (2002)\relax
\relax
\bibitem{lit-ZEUSdet}
ZEUS~Collaboration, U.~Holm (ed.), {\it The ZEUS Detector,} Status Report
  (unpublished), available on \protect\\
  \verb+http://www-zeus.desy.de/bluebook/bluebook.html+\relax
\relax
\bibitem{nim:a279:290}
N.~Harnew \etal,
\newblock Nucl.\ Inst.\ Meth.{} {\bf A~279},~290~(1989)\relax
\relax
\bibitem{npps:b32:181}
B.~Foster \etal,
\newblock Nucl.\ Phys.\ Proc.\ Suppl.{} {\bf B~32},~181~(1993)\relax
\relax
\bibitem{nim:a338:254}
B.~Foster \etal,
\newblock Nucl.\ Inst.\ Meth.{} {\bf A~338},~254~(1994)\relax
\relax
\bibitem{nim:a309:77}
M.~Derrick \etal,
\newblock Nucl.\ Inst.\ Meth.{} {\bf A~309},~77~(1991)\relax
\relax
\bibitem{nim:a309:101}
A.~Andresen \etal,
\newblock Nucl.\ Inst.\ Meth.{} {\bf A~309},~101~(1991)\relax
\relax
\bibitem{nim:a321:356}
A.~Caldwell \etal,
\newblock Nucl.\ Inst.\ Meth.{} {\bf A~321},~356~(1992)\relax
\relax
\bibitem{nim:a336:23}
A.~Bernstein \etal,
\newblock Nucl.\ Inst.\ Meth.{} {\bf A~336},~23~(1993)\relax
\relax
\bibitem{lit-FPC}
A.~Bamberger et al., Nucl. Inst. Meth. {\bf A~450}, 235 (2000)\relax
\relax
\bibitem{desy-92-066}
J.~Andruszk\'ow \etal,
\newblock Preprint \mbox{DESY-92-066}, DESY, 1992\relax
\relax
\bibitem{zfp:c63:391}
ZEUS \coll, M.~Derrick \etal,
\newblock Z.\ Phys.{} {\bf C~63},~391~(1994)\relax
\relax
\bibitem{acpp:b32:2025}
J.~Andruszk\'ow \etal,
\newblock Acta Phys.\ Pol.{} {\bf B~32},~2025~(2001)\relax
\relax
\bibitem{lit-briskin}
G.M.~Briskin, Ph.D Thesis, Report DESY-THESIS-1998-036, DESY (1998)\relax
\relax
\bibitem{epj:c11:35}
ZEUS \coll, J.~Breitweg \etal,
\newblock Eur.\ Phys.\ J.{} {\bf C~11},~35~(1999)\relax
\relax
\bibitem{lit-turcato}
M.~Turcato, Ph.D Thesis, Report DESY-THESIS-2003-039, DESY (2003)\relax
\relax
\bibitem{lit-jb}
F.~Jacquet and A.~Blondel, {\it Detection of the Charged Current Event --
  {Method}~{II},} ``Proceedings of the Study for an $ep$ Facility for
  {Europe}", U.~Amaldi (ed.), p. 391, Hamburg, Germany (1979), also in preprint
  DESY 79/48\relax
\relax
\bibitem{lit-ktalgorithm}
S.~Catani \etal,
\newblock Nucl.~Phys.{} {\bf B~406},~187~(1993)\relax
\relax
\bibitem{pr:d48:3160}
S.D.~Ellis and D.E.~Soper,
\newblock Phys.\ Rev.{} {\bf D~48},~3160~(1993)\relax
\relax
\bibitem{lit-xgamma}
ZEUS Collaboration, M.~Derrick \etal,
\newblock Phys.~Lett.{} {\bf B~348},~665~(1995)\relax
\relax
\bibitem{nim:a355:278}
W.H.~Smith \etal,
\newblock Nucl.\ Inst.\ Meth.{} {\bf A~355},~278~(1995)\relax
\relax
\bibitem{Smith:1992b}
W.H.~Smith, K.~Tokushuku and L.W.~Wiggers, ``Proc. Computing in High-Energy
  Physics (CHEP)", Annecy, France, Sept. 1992, C.~Verkerk and W.~Wojcik (eds.),
  p. 222, CERN, Geneva, Switzerland (1992), also preprint DESY-92-150B\relax
\relax
\bibitem{lit-golec}
K.~Golec-Biernat, J.~Kwiecinski and A.~Szczurek,
\newblock Phys.~Rev.{} {\bf D~56},~3995~(1997)\relax
\relax
\bibitem{lit-SK}
S.~Kagawa, Ph.D. Thesis, KEK-report 2005-12 (2006)\relax
\relax
\bibitem{lit-renner}
R.~Renner, Ph.D. Thesis, Bonn-IR-2006-13 (2006)\relax
\relax
\bibitem{cpc:86:147}
H.~Jung,
\newblock Comp.\ Phys.\ Comm.{} {\bf 86},~147~(1995)\relax
\relax
\bibitem{lit-heracles}
K.~Kwiatkowski, H.~Spiesberger and H.J.~M\"ohring,
\newblock Comp.~Phys.~Comm.{} {\bf 69},~155~(1992)\relax
\relax
\bibitem{zfp:c76:613}
H1 \coll, C.~Adloff \etal,
\newblock Z.\ Phys.{} {\bf C~76},~613~(1997)\relax
\relax
\bibitem{lit-GRV}
M.~Gl\"uck, E.~Reya and A.~Vogt,
\newblock Phys.~Rev.{} {\bf D~51},~3220~(1995)\relax
\relax
\bibitem{zfp:c37:465}
M.~Bengtsson and T.~Sj\"ostrand,
\newblock Z.\ Phys.{} {\bf C~37},~465~(1998)\relax
\relax
\bibitem{cpc:39:347}
T.~Sj\"ostrand,
\newblock Comp.\ Phys.\ Comm.{} {\bf 39},~347~(1986)\relax
\relax
\bibitem{cpc:43:367}
T.~Sj\"ostrand and M.~Bengtsson,
\newblock Comp.\ Phys.\ Comm.{} {\bf 43},~367~(1987)\relax
\relax
\bibitem{cpc:46:43}
M.~Bengtsson and T.~Sj\"ostrand,
\newblock Comp.\ Phys.\ Comm.{} {\bf 46},~43~(1987)\relax
\relax
\bibitem{cpc:82:74}
T.~Sj\"ostrand,
\newblock Comp.\ Phys.\ Comm.{} {\bf 82},~74~(1994)\relax
\relax
\bibitem{tech:cern-dd-ee-84-1}
R.~Brun et al.,
\newblock {\em {\sc geant3}},
\newblock Technical Report CERN-DD/EE/84-1, CERN, 1987\relax
\relax
\bibitem{cpc:144:104}
B.E.~Cox and J.R.~Forshaw,
\newblock Comp.\ Phys.\ Comm.{} {\bf 144},~104~(2002)\relax
\relax
\bibitem{jhep:01:010}
G.~Corcella \etal,
\newblock JHEP{} {\bf 01},~010~(2001)\relax
\relax
\bibitem{lit-webber1}
B.R.~Webber,
\newblock Preprint \mbox{hep-ph/9411384}, 1994\relax
\relax
\bibitem{lit-schmell}
M.~Schmelling,
\newblock Physica~Scripta{} {\bf 51},~683~(1995)\relax
\relax
\bibitem{manual:cern-th-7112/93}
T.~Sj\"ostrand,
\newblock {\em {\sc{Pythia} 5.7} and {\sc{Jetset} 7.4} Physics and Manual},
  1993.
\newblock CERN-TH 7112/93\relax
\relax
\bibitem{epj:c12:375}
CTEQ \coll, H.L.~Lai \etal,
\newblock Eur.\ Phys.\ J.{} {\bf C~12},~375~(2000)\relax
\relax
\bibitem{desy-95-069}
M. Kasprzak,
\newblock Preprint \mbox{DESY-95-069}, 1995\relax
\relax
\bibitem{np:b672:3}
ZEUS \coll, S.~Chekanov \etal,
\newblock Nucl.\ Phys.{} {\bf B~672},~3~(2003)\relax
\relax
\bibitem{lit-klasen}
M.~Klasen and G.~Kramer,
\newblock Eur.~Phys.~J.{} {\bf C~38},~93~(2004)\relax
\relax
\bibitem{lit-AFG}
P.~Aurenche, M.~Fontannaz and J.~P.~Guillet,
\newblock Eur.~Phys.~J.{} {\bf C~44},~395~(2005)\relax
\relax
\bibitem{lit-frix3}
S.~Frixione and S.~Ridolfi, Nucl. Phys. {\bf B~507}, 315 (1997)\relax
\relax
\end{mcbibliography}
\end{document}